\documentstyle[prl,aps,epsfig,preprint]{revtex}
\begin{document}
\tighten

\title{Off-Forward Parton Distributions}
\author{Xiangdong Ji}
\bigskip

\address{
Department of Physics \\
University of Maryland \\
College Park, Maryland 20742 \\
{~}}

\date{UMD PP\#98-092 ~~~DOE/ER/40762-144~~~ February 1998}

\maketitle

\begin{abstract}
Recently, there have been some interesting 
developments involving off-forward parton 
distributions of the nucleon, deeply virtual 
Compton scattering, and hard diffractive vector-meson 
production. These developments
are triggered by the realization that the off-forward
distributions contain information about the 
internal spin structure
of the nucleon and that diffractive 
electroproduction of vector mesons depends on these 
unconventional distributions. This paper gives a brief
overview of the recent developments. 
 
\end{abstract}
\pacs{xxxxxx}

\narrowtext
\section{introduction}

Parton distributions were first introduced 
by Feynman as the phenomenological quantities 
describing the properties of the nucleon
manifest in high-energy scattering\cite{fey}. When 
a nucleon travels near the speed of light, 
it can be viewed as consisting of a beam of 
non-interacting partons that are massless 
and collinear. Parton distributions are 
the densities of partons as a function of 
the fraction $x$ of the nucleon 
momentum they carry. With the advent of Quantum 
Chromodynamics (QCD), Feynman's parton model, 
modulo logarithms, can be justified in 
a rigorous field theoretical approach \cite{col1}.
Meanwhile, parton distributions can be expressed 
as the matrix elements of the bilocal light-ray 
operators in the ground state of the nucleon
\cite{col2}. Only in the light-like gauge, $A^+=0$,
and light-cone quantization, does their physical 
interpretation become apparent. 

When studying the physics of parton distributions 
themselves, one of course can no longer regard
the nucleon as consisting of free 
partons. Calculating these distributions
requires a detailed knowledge of the nucleon 
wave function in quark and gluon degrees of freedom,
and undoubtedly is one of the most challenging problems 
in nonperturbative QCD. At present, the only 
fundamental approach we know is the lattice 
QCD method \cite{lat}. On the phenomenological 
side, however, the unpolarized parton
distributions have been extracted with good
precision from various high-energy scattering
data taken in the last thirty years \cite{grv,msr,cteq}. 
The empirical parton distributions not only provide
the necessary input for calculating high-energy 
scattering cross sections, e.g., of top-quark 
production, but also contain valuable
information about how the nucleon is made of its
constituents.

Recently, there has been much theoretical activity 
surrounding the so-called off-forward 
parton (or non-forward, off-diagonal) 
distributions (OFPD's). The new distributions generalize 
those introduced by Feynman: They characterize 
certain properties of the nucleon exhibited in a class 
of high-energy scattering; they reflect the 
low-energy internal structure of the particle. 
However, there are also important 
differences between the off-forward and forward
distributions. The OFPD's in general cannot be 
regarded as particle densities, but rather 
their physical interpretation is given in terms
of probability amplitude. As we shall see, they
do have simple physical significance 
in light-cone coordinates (or the infinite 
momentum frame). 

The most natural appearance of the OFPD's
is in {\it non-forward} high-energy processes. 
As far as we know, such processes were 
first considered by 
Watanabe \cite{wat} and by Bartels and M. Loewe 
\cite{bar}. In Ref. \cite{wat}, a general virtual-photon 
Compton scattering off a nucleon target was studied 
in operator product expansion (OPE), an integral 
representation of the amplitude was obtained, and 
the off-forward matrix elements of twist-two 
operators were defined. The study, however, 
mentioned little about physical motivations and 
the practicality of the process; in particular,
significance of the integral representation
was obscure. In Ref. \cite{bar}, 
diffractive production of $Z^0$ and photons was
studied in perturbative QCD and the properties of the 
non-diagonal gluon ladder were investigated. However, 
no notion of an off-forward distribution of the nucleon
target was apparent in the paper. Following \cite{bar}, 
a set of off-forward evolution kernels were 
presented by Gribov, Levin, and Ryskin
\cite{gri}. These kernels were
considered as those for the scattering amplitude. 
The role of off-forward distributions in hard
diffractive processes had remained obscured in relatively
recent studies. In Ref. \cite{rys1}, diffractive 
electroproduction of $J/\psi$ was analyzed by Ryskin
in the small $x$ region. The production amplitude was related
to the usual Feynman gluon distribution. The off-forwardness
was taken into account by a phenomenological form factor.
In Refs. \cite{bro1,rys2,fra1}, Ryskin's analysis was 
generalized to light vector-meson 
production in which the meson structure is accounted 
for by the leading-twist light-cone wave function.  

To the author's knowledge, one of first examples of 
OFPD's was introduced 
by Dittes et al. \cite{dit}, and was called 
``interpolating function.'' Ironically, the 
main motivation for introducing 
such an object was not from studying any 
off-forward process; rather, it was considered 
as an example of the physical quantities whose
evolution kernel interpolates between the 
two well-known cases: the DGLAP and ERBL evolutions
\cite{dglap,efr,bro3}. 
A serious study of the OFPD in a physical process
was made by D. M\"uller et al. \cite{mul1} 
in which they showed that it contributes to the general 
two virtual-photon Compton scattering---
the same process considered earlier by Watanabe. They
also derived the evolution equations for the OFPD. 
Note that in Ref. \cite{jai} a general off-forward amplitude 
(or single-particle Green's function) was 
introduced in studying hadron helicity flip 
in hard scattering. However, because the correlation of 
two quark fields in the object is off the light cone, 
it does not have a simple parton interpretation 
as the OFPD's do.  

The recent wave of interest in the OFPD's was generated 
from two new developments. The first development
is the recognition of the importance of the 
OFPD's in describing the internal structure
of the nucleon \cite{ji1}. In studying the 
spin structure, it was found that the fractions
of the nucleon spin carried by quarks and gluons 
can be determined by the form factors of the 
corresponding energy-momentum tensors. These form
factors can be obtained from the moments of 
the OFPD's which were independently introduced 
in \cite{ji1}. In the same paper, deeply virtual
Compton scattering (DVCS) was proposed as a practical
means to measure the OFPD's. In DVCS, a highly virtual 
photon scatters off a quark in a nucleon target, 
which then emits a real photon and 
returns to form a recoil nucleon. Although 
the two-photon process studied earlier
by Watanabe and M\"uller et al. contains DVCS as its 
special kinematic limit, the latter has important 
experimental advantages and a special theoretical 
status. Indeed, DVCS is more delicate 
to treat theoretically because of an extra light-like
vector among the external momenta.

The second development is the realization that hard
diffractive electroproduction of vector mesons requires
the use of OFPD's \cite{abr,ji1}. The first
serious studies in this direction were made by Radyushkin 
\cite{rad1}, Hoodbhoy \cite{hoo1}, and Collins et al. \cite{col3}. 
In Ref. \cite{rad1}, an off-forward gluon 
distribution - most important in the 
low $x$ region - was introduced and
used to calculate vector meson production 
in deep-inelastic electron scattering. The result
is free of the assumptions made in Ref. \cite{bro1}
about the dominance of the absorptive contribution
and the use of the forward gluon distribution. 
Hoodbhoy \cite{hoo1} used the off-forward gluon distribution
to calculate heavy quarkonium production
investigated earlier by Ryskin \cite{rys1}. He obtained
a similar result as Radyushkin's.
In a comprehensive paper, Collins, Frankfurt, and Strikman 
showed that a factorization theorem
exists for general hard diffractive meson production
in deep-inelastic scattering \cite{col3}. 
The factorization theorem explicitly involves
the OFPD's. 

Recently, many works have been produced in studying 
off-forward parton distributions, deeply virtual Compton
scattering, and diffractive vector meson production. 
It is the goal of this paper to review these 
interesting developments. In section II, off-forward 
distributions are introduced from different
perspectives and various definitions in the literature 
are compared. In section III, we make a 
digression to discuss the spin structure of the nucleon.
The role of the OFPD's in this interesting subject
is exposed. In section IV, we summarize our 
present knowledge about the OFPD's at hadron 
mass scales and their evolution to high-energy scales.
In section V, we consider processes in which 
the OFPD's may actually be measured:
deeply virtual Compton scattering and hard
diffractive vector-meson production. Section VI contains
the summary and outlook.

\section{Definitions and Interpretations}

Since the off-forward distributions contain rich 
information about the nucleon structure, it is
instructive to examine their definitions and
interpretations from different angles. In this section,
we will first introduce the off-forward 
distributions from the form factors of twist-two
operators and then we explore their
significance in light-cone coordinates. 
Finally we compare the OFPD's with other
equivalent definitions in the literature.

\subsection{Elastic Form Factors of Twist-Two Operators}

From the point view of the low-energy nucleon structure, 
it is perhaps most interesting to consider
the off-forward distributions as the generating 
functions for the form factors of the so-called 
twist-two operators. 
Recall that the matrix elements of the electromagnetic
current in the equal momentum states are entirely 
determined by symmetry, whereas those in the unequal momentum
states define the (Dirac and Pauli) form factors which
contain such interesting information as charge radius
and magnetic moment of the nucleon. The following
tower of twist-two operators represents a generalization 
of the electromagnetic current
\begin{equation}
    {\cal O}^{\mu_1\cdots\mu_n}_q =  
       \overline \psi_q i\stackrel{\leftrightarrow}{\cal D}^{(\mu_1}
        \cdots  i\stackrel{\leftrightarrow}{\cal D}^{\mu_{n-1}}
        \gamma^{\mu_n)} \psi_q\ ,
\label{O1} 
\end{equation}
where all indices are symmetrized and traceless (indicated
by $(\cdots)$) and $\stackrel{\leftrightarrow}{\cal D}
= (\stackrel{\rightarrow}{\cal D} - \stackrel{\leftarrow}{\cal D})/2. $
Technically, these operators transform like $(n/2,n/2)$ 
under Lorentz transformations. Our interest in such operators
is not accidental. They appear, for instance, in the 
operator production expansion of the two electromangetic 
currents in deep-inelastic scattering \cite{iz}. 
Thus although these generalized 
currents do not couple directly to any known fundamental 
interactions, they can nonetheless be studied 
indirectly in hard scattering processes.

Since the operators for $n>1$ 
are not related to any symmetry in the QCD
lagrangian, their matrix elements between 
the equal momentum states, 
\begin{equation}
  \langle  P | {\cal O}^{\mu_1\cdots\mu_n}_q|P\rangle =  
      2a_n(\mu) P^{(\mu_1} \cdots P^{\mu_n)}\ , 
\end{equation}
contain valuable dynamical information about the
internal structure of the nucleon. In this paper, we use 
the covariant normalization $\langle P|P\rangle
= 2E(2\pi)^3\delta^3(0)$ for the nucleon state. The $\mu$ 
dependence of the above matrix elements signifies
the dependence on renormalization scale and scheme \cite{iz}. 
The quark distribution $q(x, \mu)$ introduced 
by Feynman has a simple connection to the 
above matrix elements: 
\begin{equation}
       \int^1_{-1} dx x^{n-1} q(x,\mu) = a_n(\mu) \ ,   
\end{equation}
where $q(x,\mu)$ is chosen to have support 
in $(-1,1)$. For $x>0$, $q(x, \mu)$ is the just 
density of quarks which carry the $x$ fraction of 
the parent nucleon momentum. The density of antiquarks 
is customarily denoted as $\bar q(x,\mu)$, which in 
the above notation is $-q(-x,\mu)$ \cite{col1}. 

Just like the form factors of the electromagnetic
current, additional information about nucleon structure 
can be found in the form factors of the twist-two operators
when the matrix elements are taken between the states of
unequal momenta. Using Lorentz symmetry and parity and
time reversal invariance, one can write down all possible
form factors of the spin-$n$ operator 
\begin{eqnarray}
\langle P'| O^{\mu_1\cdots \mu_n}_q |P\rangle
   &= &{\overline U}(P') \gamma^{(\mu_1} U(P) \sum_{i=0}^{[{n-1\over 2}]}
       A_{qn,2i}(t) \Delta^{\mu_2}\cdots \Delta^{\mu_{2i+1}} 
      \overline{P}^{\mu_{2i+2}}\cdots
      \overline{P}^{\mu_n)}  \nonumber \\ 
    &&  + ~  {\overline U}(P'){\sigma^{(\mu_1\alpha} 
     i\Delta_\alpha \over 2M}U(P)   \sum_{i=0}^{[{n-1\over 2}]}
     B_{qn,2i}(t) \Delta^{\mu_2}\cdots \Delta^{\mu_{2i+1}} 
      \overline{P}^{\mu_{2i+2}}\cdots
    \overline{P}^{\mu_n)} \nonumber \\
     &&  + ~ C_{qn}(t) ~{\rm Mod}(n+1,2)~{1\over M}\bar U(P') U(P) 
    \Delta^{(\mu_1} \cdots \Delta^{\mu_n)} \ , 
\label{form}
\end{eqnarray}
where ${\overline U}(P')$ and $U(P)$ are the Dirac spinors
and $Mod(n+1,2)$ is 1 when $n$ even, 0 when $n$ odd.
The four-momentum transfer is denoted by $\Delta = P'-P$ and
its invariant by $t=\Delta^2$. The average nucleon momentum 
$\overline P = (P'+P)/2$ is also a useful variable.
For $n\ge 1$, even or odd, there are $n+1$ 
form factors. $C_{qn}(t)$ is present only when $n$ is even.

As the forward matrix elements $a_n$, the above form 
factors can be used to define
a new type of parton distributions---{\it 
off-forward distributions}. 
To accomplish this, we introduce a light-like vector $n^\mu$, 
which is conjugate to $\overline P$ in the sense
that $\bar P \cdot n = 1$. Write $\overline P = 
p + (\overline M^2 /2) n $, where $\overline M^2 = M^2-t/4$
and $p$ is another light-like vector.  
Contracting both sides of Eq. (\ref{form}) with $n_{\mu_1}
\cdots n_{\mu_n}$, we have
\begin{equation}
n_{\mu_1}\cdots n_{\mu_n} \langle P'| O^{\mu_1\cdots \mu_n}_q |P\rangle
  =   {\overline U}(P') \not\!n U(P)   
     H_{qn}(\xi, t)
  + {\overline U}(P'){\sigma^{\mu\alpha} n_\mu i\Delta_\alpha 
    \over 2M}U(P) E_{qn}(\xi, t) \ , 
\end{equation}
where 
\begin{eqnarray}
   H_{qn} (\xi,t) &= &\sum_{i=0}^{[{n-1\over 2}]}
     A_{qn,2i}(t) (-2\xi)^{2i} + {\rm Mod}(n+1,2)~ C_{qn}(t) (-2\xi)^n \ , \nonumber \\
   E_{qn} (\xi,t) &= &\sum_{i=0}^{[{n-1\over 2}]}
     B_{qn,2i}(t) (-2\xi)^{2i} - {\rm Mod}(n+1,2)~ C_{qn}(t) (-2\xi)^n \ .  
\end{eqnarray}
Here we have defined a new variable $\xi = -n\cdot 
\Delta/2$. Clearly, the $\xi$
dependence of $H_{qn}$ and $E_{qn}$ helps to 
distinguish among the different form factors of the same operator.
Now one can introduce the off-forward parton distributions
(OFPD's) $H(x, \xi, t)$ and $E(x, \xi,t)$ through their moments
\begin{eqnarray}
     \int^1_{-1} dx x^{n-1} E_q(x,\xi,t) & = &E_{qn}(\xi, t) \ , \nonumber \\
    \int^1_{-1} dx x^{n-1} H_q(x,\xi,t) & = &H_{qn}(\xi, t) \ . 
\end{eqnarray}     
Since all form factors are real, the new distributions
are consequently real. Moreover, because of time-reversal 
and hermiticity, they are {\it even} 
functions of $\xi$. For simplicity, 
we have not shown here the renormalization scale dependence. 
Frequently we will also omit the $t$ dependence.

The new distributions are more complicated than the
Feynman parton distributions because of their
dependence on the momentum transfer $\Delta$. As such, the 
OFPD's contain two more scalar variables besides 
the Feynman variable $x$. The variable $t$ is the usual $t$-channel 
invariant which is always present in a form factor. 
The $\xi$ variable is a natural product of marrying the concepts 
of the Feynman distribution and form factor: The former requires
the presence of a prefered momentum $p^\mu$ along which the partons are 
predominantly moving and the latter requires a four-momentum
transfer $\Delta$; $\xi$ is just a scalar product of the two 
momenta. Some insights about the dependence of the OFPD's
on these two extra variables will be offered later in 
Section IV.A.

In QCD, there are five additional towers of twist-two operators
besides that in Eq. (\ref{O1}): 
\begin{eqnarray}
    \tilde {\cal O}^{\mu_1\cdots\mu_n}_q &= & 
       \overline \psi_q i\stackrel{\leftrightarrow}{\cal D}^{(\mu_1}
        \cdots  i\stackrel{\leftrightarrow}{\cal D}^{\mu_{n-1}}
        \gamma^{\mu_n)} \gamma_5 \psi_q\ , \nonumber \\
    {\cal O}^{\mu_1\cdots\mu_n\alpha}_{qT} &= & 
       \overline \psi_q i\stackrel{\leftrightarrow}{\cal D}^{(\mu_1}
        \cdots  i\stackrel{\leftrightarrow}{\cal D}^{\mu_{n-1}}
        \sigma^{\mu_n)\alpha} \psi_q\ , \nonumber \\
   {\cal O}^{\mu_1\cdots\mu_n}_g &= & 
       F^{(\mu_1\alpha} i\stackrel{\leftrightarrow}{\cal D}^{\mu_2}
        \cdots  i\stackrel{\leftrightarrow}{\cal D}^{\mu_{n-1}}
        F_\alpha^{~\mu_n)} \ , \nonumber \\
  \tilde {\cal O}^{\mu_1\cdots\mu_n}_g &= & 
       F^{(\mu_1\alpha} i\stackrel{\leftrightarrow}{\cal D}^{\mu_2}
        \cdots  i\stackrel{\leftrightarrow}{\cal D}^{\mu_{n-1}}
        i\tilde F_\alpha^{~\mu_n)} \ , \nonumber \\
    {\cal O}^{\mu_1\cdots\mu_n\alpha\beta}_{gT} &= & 
       F^{(\mu_1\alpha} i\stackrel{\leftrightarrow}{\cal D}^{\mu_2}
        \cdots  i\stackrel{\leftrightarrow}{\cal D}^{\mu_{n-1}}
        F^{\mu_n)\beta}\ . 
\end{eqnarray}
The corresponding off-forward distributions can be labelled by
$\Big(\tilde H_{q}(x,\xi)$, $\tilde E_{q}(x, \xi)\Big)$, 
$\Big(H_{Tq}(x,\xi)$, $E_{Tq}(x, \xi)\Big)$, 
$\Big( H_{g}(x,\xi)$, $E_{g}(x, \xi)\Big)$,
$\Big(\tilde H_{g}(x,\xi)$, $\tilde E_{g}(x, \xi)\Big)$, 
and $\Big(H_{Tg}(x,\xi)$, $E_{Tg}(x, \xi)\Big)$, 
respectively.

\subsection{OFPD's in Light-Cone Coordinates and Gauge}

The physical significance of parton distributions 
in high-energy processes is apparent only in light-cone
coordinates and light-cone gauge. To see this, 
we sum up all the local twist-two operators into
a light-cone bilocal operator and express 
the parton distributions in terms of the latter,
\begin{eqnarray}
   F_q(x, \xi, t) && = {1\over 2}\int {d\lambda\over 2\pi} e^{i\lambda x}
    \left\langle P'\left|\overline \psi_q \left(-{\lambda \over 2}n\right)
      \not \! n {\cal P}e^{-ig\int^{-\lambda/2}_{\lambda/ 2}
       d\alpha ~n\cdot A(\alpha n)} 
    \psi_q\left({\lambda \over 2}n\right) \right| P\right\rangle 
    \nonumber \\
  && = H_q(x, \xi, t)~ {1\over 2}\overline U(P')\not\! n U(P)
    + E_q(x, \xi, t)~ {1\over 2}\overline U(P') {i\sigma^{\mu\nu}
  n_\mu \Delta_\nu \over 2M} U(P) \ . 
\label{string}
\end{eqnarray}  
The light-cone bilocal operator (or light-ray
operator) arises frequently in hard scattering processes
in which partons propagate along the light-cone. In fact,
the Taylor-expansion of this operator along the 
light-cone leads us immediately 
to the twist-two operators considered in the last subsection.
The parton distributions are most naturally defined
in terms of the matrix elements of the bilocal operator.  
In this context, the Feynman $x$ is just the Fourier 
variable of the light-cone distance. The Lorentz structures 
in the second line in the above equation are independent
and complete.

In ordinary coodinates, the correlation of two quark 
fields in the light-ray operator is in both space 
and time. As such, the physical interpretation
is quite complicated. However, in light-cone coordinates,
the physical content of the bilocal operator becomes
transparent. To see this, let us recall that the light-cone
coordinates are defined as, 
\begin{eqnarray}
     x^{\pm} &=& {1\over \sqrt{2}}(x^0 \pm x^3) \ , \nonumber \\  
     x_\perp &=& (x^1, x^2) \ . 
\end{eqnarray}
In this new system of coordinates, the role of time is 
played by $x^+$. In the same fashion, introduce the 
light-cone Dirac matrices $\gamma^{\pm}$ and the 
projection operators, $P_\pm = {1\over 2} \gamma^{\mp}
\gamma^{\pm}$. A Dirac field $\psi$ can be decomposed
into a sum of $\psi_+=P_+\psi$, the independent 
degrees of freedom (``good"), and $\psi_-= P_-\psi$, 
the dependent degrees of freedom (``bad"). In light-cone
quantization, the independent components of the
Dirac field obey the following commutation relation \cite{bro2} 
\begin{equation}
    \left[\psi_{+\alpha}(x), \psi_{+\beta}^\dagger(y)
    \right]_+|_{x^+=y^+} = {1\over \sqrt{2}}
        P_{+\alpha\beta}\delta(x^--y^-)\delta^2(x_\perp
     - y_\perp)  \ , 
\end{equation}
where $[~]_+$ denotes anticommutation.
The above relation can be solved 
by a Fourier expansion 
\begin{eqnarray}
  \psi_+(x^-,x_\perp) &=& \int {dk^+ d^2\vec{k}_\perp \over 2k^+(2\pi)^3} 
     \theta(k^+) \sum_{\lambda = \pm}
      \left(b_\lambda(k^+,\vec{k}_\perp) u_\lambda(k) 
       e^{-i(x^-k^+-\vec{x}_\perp
    \cdot \vec{k}_\perp)} \right. \nonumber \\ 
  && \left. +~ d_\lambda^\dagger(k^+,\vec{k}_\perp) v_\lambda(k)  
   e^{i(x^-k^+-\vec{x}_\perp
    \cdot \vec{k}_\perp)} \right) \ ,
\label{ff}
\end{eqnarray}
and similarly for $\overline \psi_+$. The quark (antiquark)
creation and annihilation operators, $b_{\lambda k}^\dagger$ 
($d_{\lambda k}^\dagger)$ and $b_{\lambda k}$ ($d_{\lambda k})$, 
obey the commutation relation  
\begin{eqnarray}
   \left[ b_\lambda(k^+,\vec{k}_\perp), b^\dagger_{\lambda'}({k'}^+,
      \vec{k'}_\perp) \right]_+
   &=& 2k^+ \delta_{\lambda\lambda'}(2\pi)^3 \delta(k^+-k'^+)
    \delta^2(k_\perp-k_\perp') \ , 
    \nonumber \\
 \left(~\left[d_\lambda(k^+,\vec{k}_\perp) , 
           d^\dagger_{\lambda'}({k'}^+,
      \vec{k'}_\perp)
     \right]_+ \right.
   &=& \left. 2k^+ \delta_{\lambda\lambda'}(2\pi)^3 \delta(k^+-k'^+)
   \delta^2(k_\perp-k_\perp') \ . \right) 
\end{eqnarray}
In the presence of a gauge field, the above is unchanged
if the light-cone gauge, $A^+=0$, is chosen. The same 
gauge choice eliminates the gauge link in the 
light-ray operator.

\begin{figure}
\label{fig1}
\epsfig{figure=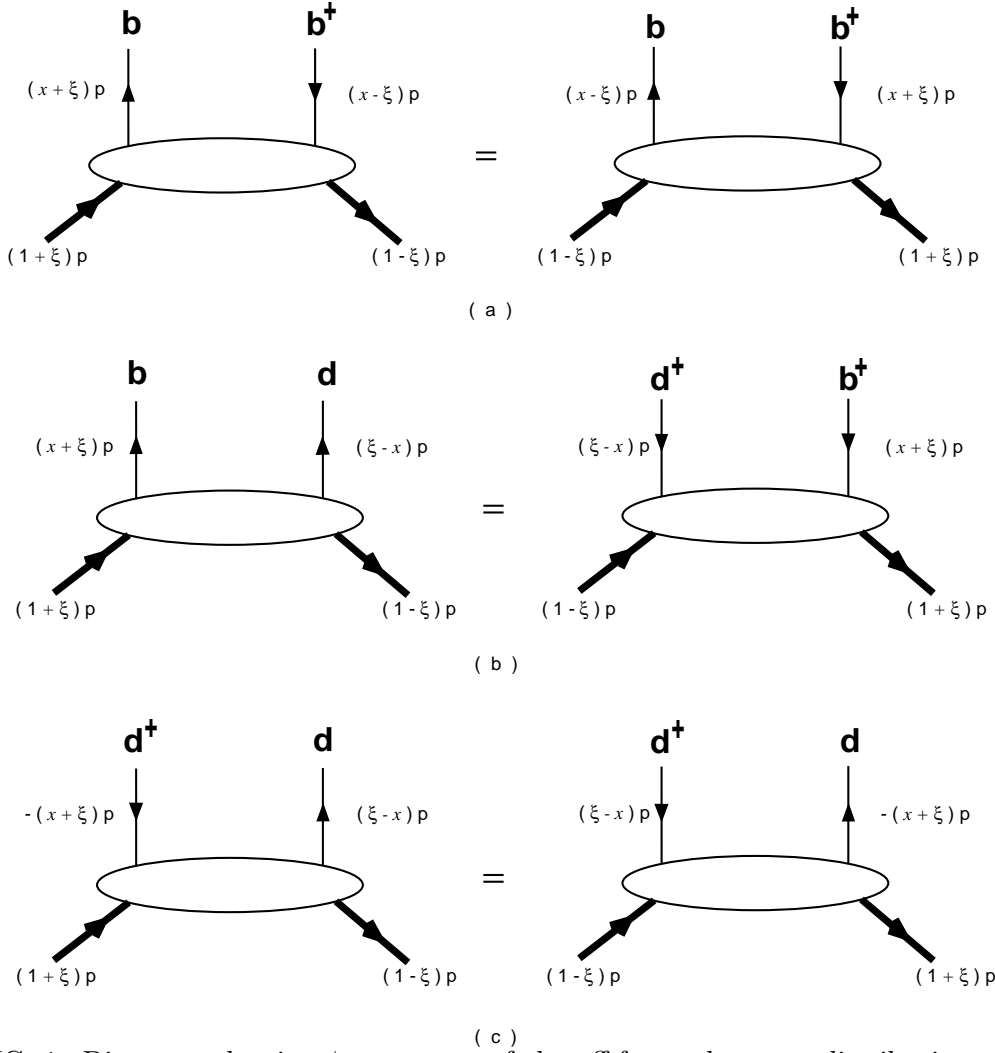,height=14.0cm}
\caption{Diagrams showing $\xi$ symmetry
of the off-forward parton distributions: a) $x>\xi$; 
b) $\xi>x>-\xi$; c) $x<-\xi$. }
\end{figure}  

Now we can look at the physical content of the off-forward
distributions in the light-cone coordinates and gauge. Since
they are even in $\xi$, we first restrict ourselves to $\xi>0$.  
Substituting the light-cone Fock expansion into Eq. (\ref{string}),
we have  
\begin{eqnarray}
      F_q(x,\xi) &= & {1\over 2p^+V}
          \int {d^2k_\perp \over 2\sqrt{|x^2-\xi^2|} (2\pi)^3}
         \sum_\lambda \nonumber \\
       &\times& \left\{ \begin{array}{ll}
         \left\langle P'\left|b^\dagger_\lambda\left((x-\xi)p^+, 
         \vec{k}_\perp+\vec{\Delta}_\perp \right) 
        b_\lambda\left((x+\xi)p^+, \vec{k}_\perp\right)\right|
        P\right\rangle\ , 
     & {\rm for}~x> \xi  \\  
       \left \langle P'\left|d_\lambda\left((-x+\xi)p^+, 
         -\vec{k}_\perp-\vec{\Delta}_\perp \right)
        b_{-\lambda}\left((x+\xi)p^+, \vec{k}_\perp\right)\right|P
       \right\rangle \ ,
      & {\rm for}~\xi > x> -\xi \\
        -\left\langle P'\left| d^\dagger_\lambda\left((-x-\xi)p^+, 
         \vec{k}_\perp+\vec{\Delta}_\perp \right)
        d_\lambda\left((-x+\xi)p^+, \vec{k}_\perp\right)\right|
         P\right\rangle \ ,  
     & {\rm for}~x< -\xi \end{array} \right.
\label{fock}
\end{eqnarray}   
where $V$ is a volume factor.
The distribution has different physical interpretations
in the three different regions. In the region $x>\xi$, it is the
amplitude for taking a quark of momentum $k$ out 
of the nucleon, changing its momentum to $k+\Delta$,  
and inserting it back to form a recoiled nucleon. 
In the region $\xi> x> -\xi$, it is the amplitude
for taking out a quark and antiquark pair with 
momentum $-\Delta$. Finally, 
in the region $x<-\xi$, 
we have the same situation as in the first, except
the quark is replaced by an antiquark. The first and third 
regions are similar to those present in 
the ordinary parton distributions and hence will be
called the DGLAP region. The middle region is similar to 
that in a meson amplitude and hence will be called 
the ERBL region. 

The $\xi$ symmetry is an important property of the
OFPD's. To illustrate the parton content of this symmetry,
one can work out the equivalent of Eq. (\ref{fock})
by assuming $\xi$ is negative. The result is best shown
in Fig. 1, where the quark and antiquark creations and
annihilations are indicated explicitly. 

From Eq. (\ref{fock}), the off-forward 
distribution is very much like a quark-nucleon 
scattering amplitude. Use ${\cal A}_{Hh,H'h'}$ 
to label the helicity amplitudes, where $H,H'$
are the helicities of the initial and final nucleons
and $h,h'$ are those of the created and annihilated 
quarks. Time-reversal and parity invariance 
provide the constraints ${\cal A}_{Hh,H'h'}
= {\cal A}_{H'h',Hh}$ and ${\cal A}_{Hh,H'h'} 
= {\cal A}_{-H-h,-H'-h'}$. Therefore, there are
total of six independent helicity amplitudes which
imply four additional twist-two off-forward
distributions \cite{hoo2}
\begin{eqnarray}
   \tilde F_q(x, \xi) && = {1\over 2}\int 
     {d\lambda\over 2\pi} e^{i\lambda x}
    \left\langle P'\left|\overline \psi_q \left(-{\lambda \over 2}n\right)
       \not \! n\gamma_5 {\cal P}e^{-ig\int^{-\lambda/2}_{\lambda/ 2}
       d\alpha ~n\cdot A(\alpha n)} 
    \psi_q\left({\lambda \over 2}n\right) \right| P\right\rangle  \nonumber \\
  && = \tilde H_q(x, \xi, t)~ {1\over 2}\overline U(P')\not\! n \gamma_5 U(P)
    + \tilde E_q(x, \xi, t)~ {1\over 2} \overline U(P') {\gamma_5\Delta \cdot n\over 2M} U(P) \ . 
     \nonumber \\
   F_{Tq}(x, \xi) && = {1\over 2}\int {d\lambda\over 2\pi} e^{i\lambda x}
    \left \langle P'\left|\overline \psi_q \left(-{\lambda \over 2}n\right)
       n_\mu\sigma^{\mu\perp} {\cal P}e^{-ig\int^{-\lambda/2}_{\lambda/ 2}
       d\alpha ~n\cdot A(\alpha n)} 
    \psi_q\left({\lambda \over 2}n\right) \right| P\right\rangle  \nonumber \\
  && = H_{Tq}(x, \xi)~ {1\over 2}\overline U(P')n_\mu\sigma^{\mu\perp} U(P)
    + E_{Tq}(x, \xi)~ {1\over 2} \overline U(P') {n_\mu \gamma^{[\mu}
 i\Delta^{\perp]} \over M} U(P) \ . 
\end{eqnarray}  
Their parton interpretation is similar to that of $F_q(x,\xi)$, 
except that for $\tilde F_q(x,\xi)$ there is a sign depending on 
the quark helicity and for $F_{Tq}(x,\xi)$ there is a quark 
helicity flip \cite{jaf1}.

Off-forward gluon distributions can be introduced 
in the similar fashion. Consider first
the gluon-nucleon helicity amplitudes. Symmetries
limit the number of independent ones
to six, and hence there are six leading-twist 
gluon distributions. Here we quote only the 
helicity-independent ones: 
\begin{eqnarray}
F_G(x, \xi) & =& {1\over 2x} \int {d\lambda\over 2\pi}
    e^{i\lambda x}
    \left\langle P'\left|F^{\mu\alpha}\left(-{\lambda\over 2}n
\right)  F_{\alpha}^{~\nu}\left({\lambda\over 2}n\right)\right|P
\right\rangle  n_\mu n_\nu  \nonumber \\
   &=& H_g(x, \xi)~ {1\over 2}\overline U(P')\not\! n U(P)
    + E_g(x, \xi)~ {1\over 2}\overline U(P') {i\sigma^{\mu\nu}
  n_\mu \Delta_\nu \over 2M} U(P) \ ,  
\end{eqnarray}
where the gauge link is not shown. Because of the Bose
symmetry, $F_G(x,\xi) = -F_G(-x, \xi)$. 
The interested reader can find the other twist-two
gluon distributions in \cite{hoo2}. 

In light-front coordinates, only $A_\perp$ are
the independent degrees of freedom. They obey the commutation
relation \cite{bro2}
\begin{equation}
   [A_i(x), A_j(y)]|_{x^+=y^+} = {1\over 2} \delta_{ij}
     \delta^2(x_\perp-y_\perp) \epsilon(x^--y^-)
\end{equation}
where $\epsilon(a-b) = \theta(a-b)-\theta(b-a)$. The
commutation relation can be solved by the following 
expansion 
\begin{eqnarray}
    A_i(x^-,x_\perp) & = & \int {dk^+ d^2\vec{k}_\perp 
     \over 2k^+(2\pi)^3} 
      \theta(k^+)\sum_{\lambda = \pm}
      \left(a_\lambda(k^+,\vec{k}_\perp)
   \epsilon_i(\lambda,k) e^{-i(x^-k^+-\vec{x_\perp}
    \cdot \vec{k_\perp})} \right.
                \nonumber \\ &&~~~~~
  \left. +a_\lambda^\dagger(k^+,\vec{k}_\perp) 
    \epsilon^*_i(\lambda, k)  
   e^{i(x^-k^+-\vec{x_\perp}
    \cdot \vec{k_\perp})} \right) \ ,
\end{eqnarray}
where $a^\dagger_\lambda(k)$ and $a_\lambda(k)$
are the gluon creation and annihilation operators.
Using this expansion, the physical content of the 
gluon distributions can be worked out immediately
\begin{eqnarray}
      F_G(x,\xi) &= & {1\over 4p^+V}
          \int {d^2k_\perp \over 2\sqrt{|x^2-\xi^2|} (2\pi)^3}
         \sum_\lambda \nonumber \\
       &\times& \left\{ \begin{array}{ll}
         \left\langle P'\left|a^\dagger_\lambda\left((x-\xi)p^+, 
         \vec{k}_\perp+\vec{\Delta}_\perp \right) 
        a_\lambda\left((x+\xi)p^+, 
     \vec{k}_\perp\right)\right|P\right\rangle\ , 
     & {\rm for}~x> \xi   \\
       \left \langle P'\left|a_\lambda\left((-x+\xi)p^+, 
         -\vec{k}_\perp-\vec{\Delta}_\perp \right)
        a_{-\lambda}\left((x+\xi)p^+, \vec{k}_\perp\right)\right|P
       \right\rangle \ ,
     & {\rm for}~\xi>x>0 \end{array} \right.
\end{eqnarray}
In the region $x>\xi$, it is the amplitude for taking
a gluon of momentum $k$ out of the nucleon, changing its
momentum to $k+\Delta$, and inserting it back to form a recoiled
nucleon. In the region $\xi>x>0$, it is the amplitude for
taking out a gluon pair with momentum $-\Delta$. 

In an actual calculation of the distributions, the vacuum 
subtraction in matrix elements is always implied. As such, it 
is convenient to insert a $T$ product between the two fields
at light-cone separations. The difference between a 
$T$ product and an ordinary product
is just a constant
\begin{equation}
    T\psi^\dagger_+(0) \psi_+(\lambda n)
   = \psi^\dagger_+(0) \psi_+(\lambda n)
   - \theta(\lambda n^0)\left[\psi^\dagger_+(0),
   \psi_+(\lambda n)\right]_+ \ . 
\end{equation}
The constant does not contribute to any physical matrix element. 
Therefore, a light-cone $T$ product
contains only dispersive contributions.

\subsection{Other Notations and Definitions}

The notion of OFPD's we discussed above are essentially that of
Refs. \cite{mul1} and \cite{ji1}. The variables $x$ and 
$\xi$ here are the same as the variables $t$ and $\tau$ 
in \cite{mul1}. Other names (non-forward distributions, 
off-diagonal distributions, double distributions) 
and notations exist in the literature.

Radyushkin introduced the {\it non-forward distributions} 
${\cal F}_\zeta(X)$ in Refs. \cite{rad1,rad2}. To see 
their relationship to the OFPD's, one can write 
Eq. (\ref{string}) for an arbitrary light-like $n$ 
\begin{eqnarray}
  && ~~ \int {d\lambda\over 2\pi} e^{i\lambda k\cdot n}
    \left\langle P'\left|\overline \psi_q\left(-{\lambda \over 2}n\right)
       \not\! n {\cal P} e^{-ig\int^{-\lambda/2}_{\lambda/2}
       d\alpha~ n\cdot A(\alpha n)}
    \psi_q\left({\lambda \over 2}n\right) \right| P\right\rangle  \nonumber \\
  && = H_q\left({n\cdot k\over n\cdot \overline P}, 
         -{n\cdot \Delta \over 2n\cdot \overline P}\right) {1\over n\cdot
\overline  P} \overline U(P')\not\! n U(P) \nonumber \\
    && + ~E_q\left({n\cdot k\over n\cdot \overline P},
         -{n\cdot \Delta \over 2n\cdot \overline P}\right) {1\over n\cdot
\overline  P} \overline U(P') {i\sigma^{\mu\nu}
  n_\mu \Delta_\nu \over 2M} U(P) \ . 
\end{eqnarray} 
One can also shift the argument of the string operator 
using Heisenberg's equations of motion.     
Now if one chooses 
\begin{equation}
    n\cdot P = 1,~~ X = k\cdot n,  ~~\zeta = -\Delta \cdot n \ , 
\end{equation}
the variables are related according to 
\begin{eqnarray}
      x &=& {X-\zeta/2 \over 1-\zeta/2}, ~~~~ \xi = {\zeta/2\over  
     1-\zeta/2} \ ; \nonumber \\
      X & = & {x+ \xi\over 1+\xi}, ~~~~ \zeta = {2\xi\over 1+\xi}\ .
\end{eqnarray} 
The matrix element of the string operator becomes
\begin{eqnarray}
  &&  \int {d\lambda\over 2\pi} e^{i\lambda X}
    \left\langle P\left|\overline \psi_q(0) \not\! n {\cal P}
       e^{-ig\int^0_{\lambda/2}
       d\alpha ~n\cdot A(\alpha n)}
    \psi_q\left(\lambda n\right) \right| P\right
     \rangle = F_q(x, \xi,t)  \nonumber \\
  && = (1+\xi)H_q(x,\xi) \overline U(P')\not\! n U(P) 
     +~(1+\xi)E_q(x,\xi) 
      \overline U(P') {i\sigma^{\mu\nu}
  n_\mu \Delta_\nu \over 2M} U(P) \ . 
\end{eqnarray}     
Compared with Eqs. (9.1) and (9.2) in Ref. \cite{rad2}, we have
\begin{equation}
    (1+\xi) H_q(x, \xi) = {\cal F}_\zeta^q(X)\theta(0<X<1)
          - {\cal F}_\zeta^{\bar q}(\zeta-X)\theta(-1+\zeta 
         \le X \le \zeta) \ . 
\end{equation}
The advantages of $H(x, \xi)$ over ${\cal F}_\zeta(X)$ are:
1) it treats the initial and final nucleon symmetrically
and thus $\xi$ symmetry is obvious; 
2) it has a simple connection with local hermitian
operators. However, the $X$ variable in ${\cal F}_\zeta(X)$ has the 
advantage of measuring one of the initial parton momenta according
to the longitudinal momentum of the initial nucleon, analogous
to the forward case. 

Collins et al. introduced the 
off-diagonal distributions $f(x_1,x_2,t)$ \cite{col3}.
They also chose the $n$ vector conjugate 
to the initial state momentum. The relations of 
their variables to ours are,
\begin{eqnarray}
    x_1 &=& {x+\xi\over 1+\xi}, ~~~~ x_2 = {x-\xi\over 1+\xi} \ ;
    \nonumber \\  
    x &=& {(x_1+x_2)/2\over 1-(x_1-x_2)/2},~~\xi = {(x_1-x_2)/2
     \over 1-(x_1-x_2)/2} \ . 
\label{x2}
\end{eqnarray}
Here $x_1$ is the same as Radyuskin's $X$.
The off-diagonal distributions are related to the OFPD's by 
\begin{eqnarray}
       f_{q/p}(x_1, x_2) 
      = F_q(x, \xi) . 
\end{eqnarray}
On the other hand, for gluons, one has 
\begin{eqnarray}
      x_1 x_2 f_{g/p}(x_1,x_2, t)
    = 2xF_g(x,\xi)\  . 
\end{eqnarray}
Again, the $\xi$ symmetry is not obvious in 
these off-diagonal distributions.

Radyushkin also introduced what he called double 
distributions $F(x,y)$, where $x$ and $y$ are 
the fractions of the initial nucleon 
momentum $P$ and the momentum transfer $\Delta$
carried by the active parton \cite{rad3}. It is 
interesting from the point of view of parametrizing 
the momentum flow out of a hadron vertex 
in a $\xi$-independent way. However, 
the physics of a double distribution
is equivalent to that of the corresponding off-forward 
distribution. 

\section{OFPD's and The Spin Structure of the Nucleon}

One of the principal motivations for  
studying off-forward distributions
is that they provide a new class of observables
for the internal structure of the nucleon. 
We believe that the physics
of strong interactions is described by the 
fundamental theory--quantum 
chromodynamics (QCD). However, QCD in the 
low-energy region is strongly 
interacting and highly relativistic, involving
an infinite number of degrees of freedom.  
No theory of this type has ever been 
solved before with a clear physical insight. 
The only available tool now for 
solving QCD bound states is lattice QCD \cite{lat}. 
At this stage, the approach has not yet been developed
to a degree of sophistication so that
one can reliably calculate fundamental observables 
with respectable precisions.  Therefore, any 
experimental information on the structure of the nucleon 
is valuable in helping to understand how
nature constructed this particle of extreme
importance to our existence.

Since Gell-Mann and Zweig's quark model, many
nucleon models have been proposed to correlate
the experimental data from 
low and high energy probes. Although these models 
are not precisely QCD, they have been quite successful
phenomenologically. However, the ultimate
question confronting models is  
how the degrees of freedom used 
in them are related to those
in QCD. This question is of high importance
because experimental probes do not couple 
to model degrees of freedom. In practice, simple 
assumptions have been
used in comparing model predictions and data, e.g.,
the constituent quarks are identified as QCD quarks at 
low-energy scales although a precise notation 
of the latter is unclear. These assumptions have  
been seriously challenged in 1987 when the European 
Muon Collaboration (EMC) measured, with an unprecedented 
precision, the proton's spin-dependent
structure function $G_1(Q^2,\nu)$ in polarized deep-inelastic 
scattering. Combining their data with the hyperon 
beta decay rates, augmented with the assumption 
of the flavor SU(3) symmetry, EMC extracted the fraction 
of the nucleon spin carried in the spin of quarks: 
\begin{equation}
     \Delta \Sigma \left(\overline Q^2\!=\!10 {\rm GeV}^2\right) 
    = 0.12\pm 0.17 \ . 
\end{equation}
This result is in flat contradiction with 
the cherished quark model prediction $\Delta \Sigma=1$. 
This so-called ``spin crisis" has touched off
an explosive activity in both theoretical and
experimental communities.  After ten more years
of careful experimental studies, the original EMC
result has essentially been confirmed, although
the small $x$ contribution to $\Delta \Sigma$ remains 
largely uncertain. A recent global analysis of the data
can be found, for instance, in Ref. \cite{alt1}. 

Many theoretical ``solutions" to 
the ``spin crisis" have been proposed in the literature. 
Apart from doubts about the experimental data
and the interpretation of them, there have been
serious attempts in explaining the EMC
result from the fundamental theory.  For instance, 
lattice QCD calculations have produced numbers close 
to the EMC result \cite{fuk,don,goc}, although 
the systematic effects of quenched approximation, 
finite lattice sizes, and large quark masses remain 
uncertain. Despite these, more fundamental approaches 
have not yet offered a complete picture about the spin 
structure of the nucleon. As to the naive 
quark model prediction, the key question is what has been 
alluded to abovel: namely, whether the quark model 
result can be directly compared with the experimental 
data, or whether the constituent 
quarks are the same as the QCD quarks probed 
in high-energy scattering. No convincing argument 
has been offered towards a positive answer. 

More recently, progress has been made in attempting
to understand how QCD builds up the nucleon spin.
The progress potentially has an important
impact on other theoretical studies, and ultimately 
on our basic knowledge of the internal structure
of the nucleon. On the one hand, it suggests 
a thorough examination of the spin structure
in a fundamental way. On the other hand,
they called for understanding the role  
of the gluon degrees of freedom in 
model building. As it turns out, a key 
element in unraveling the nucleon spin is 
the off-forward parton distributions.  

The history began with identifying the sources of the QCD 
angular momentum. Jaffe and Manohar wrote down the complete
angular momentum operator in a free-field-theory form which 
contains naturally the quark and gluon spin and orbital 
contributions \cite{jaf2}. The importance of 
the quark orbital angular momentum in QCD 
has been recognized before by 
Sehgal \cite{seh} and Ratcliffe \cite{rat} in different
contexts. The EMC result indicates, among others, that
a large fraction of the nucleon spin is carried by 
other sources of angular momentum. In order
to associate these contributions to possible 
observables, the QCD angular momentum operator
has been rewritten in a gauge-invariant 
form \cite{ji1}, 
\begin{equation}
    \vec{J}_{\rm QCD} = \vec{J}_{q} + \vec{J}_g \ , 
\end{equation}
where
\begin{eqnarray}
     \vec{J}_q &=& \int d^3x ~\vec{x} \times \vec{T}_q \nonumber \\
                 &=& \int d^3x ~\left[ \psi^\dagger 
     {\vec{\Sigma}\over 2}\psi + \psi^\dagger \vec{x}\times 
          (-i\vec{D})\psi\right]
     \ ,  \nonumber \\
     \vec{J}_g &=& \int d^3x ~\vec{x} \times (\vec{E} \times \vec{B}) \ . 
\label{ang}
\end{eqnarray}
The quark and gluon parts of the angular momentum 
are generated from the quark and gluon 
momentum densities $\vec {T}_q$ and $\vec{E}\times \vec{B}$, 
respectively. $\vec{\Sigma}$ is
the Dirac spin-matrix and the corresponding term 
is clearly the quark spin contribution. $\vec{D}= \vec{\partial}+
ig\vec{A}$ is the covariant derivative and the
associated term is the gauge-invariant quark
orbital contribution. 

With the above expression, one can easily construct
a sum rule for the spin of the nucleon. Consider
a nucleon moving in the $z$-direction, and polarized 
in the helicity eigenstate $\lambda = 1/2$. The total
helicity can be evaluated as an expectation value of
$J_z$ in the nucleon state
\begin{equation}
        {1\over 2} = {1\over 2}\Delta \Sigma (\mu) 
    + L_q(\mu) + J_q(\mu) \ , 
\end{equation}
where the three terms denote the matrix elements
of three parts of the angular momentum operator
in Eq. (\ref{ang}). The physical significance 
of each term is obvious, modulo the scale 
and scheme dependence indicated by $\mu$. The scale
dependence in $\Sigma(\mu)$ is generated from
the U(1) axial anomaly, and there have been attempts
to remove the scale dependence by subtracting 
a gluon contribution \cite{alt2,car}. Unfortunately, such
a subtraction is by no means unique. Here we adopt
the standard definition of $\Delta \Sigma (\mu)$
as the matrix element of the multiplicatively
renormalized quark spin operator. Note that
the individual term in the above equation is
independent of the momentum of the nucleon. 
In particular, it applies when the nucleon
is travelling with the speed of light (the infinite
momentum frame) \cite{ji2}.

The scale dependence of the quark and gluon 
contributions can be calculated in perturbative QCD. 
By studying renormalization of the nonlocal
operators, one can show \cite{ji3,ji1}
\begin{equation}
{\partial \over \partial \ln \mu^2}
  \left(\begin{array}{c}
         J_q(\mu) \\
          J_g(\mu)
    \end{array} \right)
   = {\alpha_s(\mu)\over 2\pi}
    {1\over 9}\left( \begin{array}{rr}
        -16 & 3n_F  \\
        16 & -3n_F  \\
      \end{array} \right)
        \left( \begin{array}{c}  
           J_q(\mu) \\
            J_g(\mu)
        \end{array} \right) \ .
\end{equation}
As $\mu\rightarrow \infty$, there exists a fixed point
solution
\begin{eqnarray}
       J_q(\infty) &=& {1\over 2} {3n_f\over 16 + 3n_f} \ ,  \nonumber
\\
       J_g(\infty) &=& {1\over 2} {16\over 16 + 3n_f} \ .
\end{eqnarray}
Thus as the nucleon is probed at an infinitely small distance
scale, approximately one-half of the spin is carried by   
gluons. A similar result has been obtained by Gross and
Wilczek in 1974 for the quark and gluon contributions to
the momentum of the nucleon \cite{gro}. Strictly speaking,
these results reveal little about the nonperturbative
structure of bound states. However, experimentally
it was found that about half of the nucleon momentum is
carried by gluons already at quite low energy scales 
\cite{cteq}. Thus the gluon degrees of freedom not only
play a key role in perturbative QCD, but also are a major
component of nonperturbative states, as they should be.
An interesting question is then, how much of the nucleon 
spin is carried by the gluons at low energy scales? 
A solid answer from the fundamental theory is not 
yet available. Recently, Balitsky and this author have
made an estimate using the QCD sum rule approach
\cite{bal1}. We found
\begin{equation}
    J_g(\mu\sim 1 {\rm GeV}) \simeq {8\over 9} {e<\bar u\sigma Gu>
     <\bar uu> \over M_{1^{-+}}^2\lambda_N^2}
\end{equation}
which yields approximately
0.25. If this calculation indicates
anything of the truth, the spin structure of the nucleon
roughly looks like 
\begin{equation}
       {1\over 2} = 0.10({\rm from~} {1\over2}\Delta \Sigma)
        + 0.15({\rm from~} L_q) + 0.25({\rm from~} J_g) \ .
\end{equation}
A quark model calculation of $J_g$ has also been made
recently by Barone et al. \cite{baro}. The result also indicates
$J_g \sim 0.25$ at low-energy scales. 
                                               
By examining carefully the definition of $J_{q,g}$
\begin{equation}
      J_{q,g}(\mu) = \left\langle P{1\over 2} \left|
         \int d^3x (\vec{x}\times \vec{T}_{q,g})^z
 \right|P{1\over 2}\right\rangle \ ,
\label{matrix}
\end{equation}
one realizes that they can be extracted from the
form factors of the quark and gluon parts of
the QCD energy-momentum tensor $T^{\mu\nu}_{q,g}$.
Specializing Eq. (\ref{form}) to ($n=2$)
\begin{equation}     
      \langle P'| T_{q,g}^{\mu\nu} |P\rangle
       = \bar U(P') \Big[A_{q,g}(t)
       \gamma^{(\mu} \bar P^{\nu)} +
   B_{q,g}(t) \bar P^{(\mu} i\sigma^{\nu)\alpha}\Delta_\alpha/2M
  +  C_{q,g}(t)\Delta^{(\mu} \Delta^{\nu)}/M \Big] U(P)\ .
\end{equation}
Taking the forward limit in the $\mu=0$ component and integrating
over 3-space, one finds that $A_{q,g}(0)$ give
the momentum fractions of the nucleon carried by
quarks and gluons ($A_q(0)+A_g(0)= 1$).
On the other hand, substituting
the above into the nucleon matrix element of Eq. (\ref{matrix}),
one finds \cite{ji1}
\begin{eqnarray}
      J_{q, g} = {1\over 2} \left[A_{q,g}(0) + B_{q,g}(0)\right] \ .
\end{eqnarray}
Therefore, the matrix elements of the energy-momentum
tensor provide the fractions of
the nucleon spin carried by quarks and gluons.
There is an analogy for this. If one knows the Dirac and Pauli
form factors of the electromagnetic current, 
$F_1(Q^2)$ and $F_2(Q^2)$,
the magnetic moment of the nucleon, defined as
the matrix element of (1/2)$\int d^3x (\vec{x} \times \vec{j})^z$,
is $F_1(0) +F_2(0)$.

Since the quark and gluon energy-momentum tensors
are just the twist-two, spin-two, parton helicity-independent
operators, we immediately have the following
sum rule from the off-forward distributions
\begin{eqnarray}
     \int^1_{-1} dx x [H_q(x, \xi, t) +
       E_q(x, \xi, t) ]
     = A_q(t) + B_q(t) \ ,
\end{eqnarray}
where the $\xi$ dependence, or $C_q(t)$
contamination, drops out. Extrapolating the sum rule
to $t=0$, the total quark (and hence quark orbital)
contribution to the nucleon spin is obtained.
A similar sum rule exists for gluons.

Thus a deep understanding of the spin structure of the 
nucleon may be achieved by measuring the OFPD's
from high-energy experiments.

\section{What do we know about OFPD's?}

Comparing with the ordinary parton distributions,
off-forward distributions depend on 
two additional variables: the $t$-channel 
longitudinal momentum fraction $\xi$ and 
the invariant mass $t$.  Because of these, 
the OFPD's contain much more information 
and are necessarily more difficult to 
model in practice. 
At present, there is little experimental
data to constrain these distributions. 
In this Section, we will describe
a few of the recent studies of the distributions 
at low-energy scales where they 
reflect directly the nonperturbative 
structure of the nucleon. We will also 
discuss their evolution equations which have 
been computed to leading logarithmic 
order in perturbation theory. Some qualitative 
features of the evolution to an asymptotically large
energy scale will be highlighted.

\subsection{OFPD's at Hadron-Mass Scale}

A few rigorous results about the OFPD's
are known. First of all, in the limit of 
$\xi\rightarrow 0$ and $t \rightarrow 0$, 
they reduce to the ordinary parton
distributions. For instance, 
\begin{eqnarray}
     H_q(x, 0, 0) &=& q(x) \ ,  \nonumber \\
     \tilde H_q(x, 0, 0) &=& \Delta q(x) \ , 
\end{eqnarray}
where $q(x)$ and $\Delta q(x)$ are the unpolarized and
polarized quark densities.  Similar equations hold
for gluon distributions. For practical purposes, 
in the kinematic region where
\begin{equation}
       \sqrt{|t|}<\!\!< M_N~~~ {\rm and} ~~~ \xi<\!\!<x
\end{equation}
an off-forward distribution may be approximated 
by the corresponding forward one. The first condition, 
$\sqrt{|t|}<\!\!< M_N$, is crucial - otherwise 
there is a significant form-factor
suppression which cannot be neglected at any $x$ and $\xi$. 
For a given $t$, $\xi$ is restricted to 
\begin{equation}
    |\xi| < \sqrt{-t/(M^2-t/4)} \ . 
\label{gs}
\end{equation} 
Therefore, when $\sqrt{|t|}$ is small, $\xi$ is 
automatically limited and there is in fact 
a large region of $x$ where the forward approximation holds. 

The first moments of the off-forward distributions
are constrained by the form factors of 
the electromagnetic and axial currents. Indeed,
by integrating over $x$, we have \cite{ji1}
\begin{eqnarray}
     \int^1_{-1} dx H_q(x, \xi, t) &=& F_1^q(t) \ , \nonumber \\ 
     \int^1_{-1} dx E_q(x, \xi, t) &=& F_2^q(t) \ , \nonumber  \\
     \int^1_{-1} dx \tilde H_q(x, \xi, t) &=& G_A^q(t) \ , \nonumber \\ 
     \int^1_{-1} dx \tilde E_q(x, \xi, t) &=& G_P^q(t) \ .
\end{eqnarray}
The $t$ dependence of the form factors are 
characterized by hadron mass scales. Therefore, 
it is reasonable to speculate that similar 
mass scales control the $t$ dependence of 
the off-forward distributions. 
In the small $x$ region, it has been argued that the 
$t$ dependence corresponds to 
a mass scale $4m_\pi^2$ or something close to 
the inverse slope of the elastic $PP$ cross section, 
$1/B = 0.1$ GeV$^2$ \cite{mar}. 

Martin and Ryskin made a very important observation that 
one can obtain an upper bound on the off-forward 
distributions by using the Schwarz inequality \cite{mar}.
For instance, consider the quark distribution 
in Eq. (\ref{fock}) and take a ket $|\psi\rangle = 
b_\lambda\left((x+\xi)p^+,
\vec{k}_\perp\right)|P\rangle/\sqrt{x+\xi} 
- b_\lambda\left((x-\xi)p^+,\vec{k}_\perp
 +\vec{\Delta}_\perp\right)|P'\rangle/\sqrt{x-\xi}$. 
Since $\langle \psi|\psi\rangle\ge 0$, we have the 
following constraint 
\begin{equation}
    F_q(x, \xi) \le {1\over 2} \left( q(x_1) 
   + q(x_2) \right) \ , 
\end{equation}
for $x>\xi$, where $x_1 = (x+\xi)/(1+\xi)$
and $x_2 = (x-\xi)/(1-\xi)$ which is different from that 
in Eq. (\ref{x2}). Similarly, one has
\begin{equation}
    -F_q(x, \xi) \le {1\over 2} \left( \bar q(-x_1) + \bar 
      q(-x_2) \right) \ , 
\end{equation}    
for $x<-\xi$. In the middle region, however, 
the vacuum subtraction ruins
a useful constraint. 
An upper bound for the off-forward gluon 
distribution can be derived in a similar way:  
\begin{equation}
    2xF_g(x, \xi) \le  {1\over 2}\left(x_1 g(x_1) + x_2g(x_2)\right) \ , 
\label{gmr}
\end{equation}
for $x>\xi$. Because of the different
definitions of $x_2$, the above inequality 
is slightly different from Martin and Ryskin's. However, 
they are practically the same in the small $x$ region. 
It must be cautioned that the above inequalities
are derived by disregarding renormalization 
of the distributions, and hence are correct 
in the leading-logarithmic approximation only.

A first model calculation of the quark off-forward
distributions was reported in Ref. \cite{ji4}. 
The calculation was done in the MIT bag model, whose
parameters are adjusted so that the electromagnetic
form factor and the Feynman parton distributions
are best produced. The shape of the distributions 
as a function of $x$ are rather similar 
at different $t$ and $\xi$: peaked around $x=0.2$
$\sim$ 0.4, and tapering off rather quickly as 
$x\rightarrow 1$, 0 and negative, where the calculation
cannot be trusted. The $t$ dependence of the energy-momentum 
form factors is controlled by 
a mass parameter between 0.5 and 1 GeV$^2$.
The $\xi$ dependence of the distributions 
is surprisingly weak; a clear physical 
reason is unknown. 

The off-forward quark distributions were also studied
in the chiral quark-soliton model by Petrov et al. \cite{pet}. 
In contrast to the bag model results, the chiral soliton
model yields a rather strong $\xi$ dependence of the
off-forward distribution. The model also 
predicts qualitatively different behaviors in the regions
$|x|>\xi$ and $|x|<\xi$, with discontinuities at $x=\pm \xi$. 
Although the discontinuities are believed to be an 
artifact of the calculation, the qualitatively
different features in different regions are in line with 
the interpretation of the OFPD's. From Eq. (\ref{fock}), 
we understand that the off-forward
distributions have similar physics as the forward 
distributions in $|x|>\xi$. When $|x|<\xi$, however, 
the distributions reflect $t$-channel exchanges of 
meson and glueball quantum numbers.  Therefore, a 
physics-oriented modelling in this region may consider 
the convolution of the meson cloud of a nucleon with 
the leading light-cone wave function of the mesons.

Models cannot be used yet to calculate off-forward 
gluon distributions in the small $x$ region,
which are phenomenologically important \cite{rys1,bro1,rys2,fra1}. 
It is believed in the literature 
that the distributions in this region and negligible 
$t$ are independent of 
$\xi$ \cite{abr,fra1}. As we mentioned earlier, this 
might be correct as long as $\xi\ll x$. For $\xi\sim x$, however, 
the best argument so far is offered by 
Martin and Ryskin \cite{mar}. From the diffractive 
scattering phenomenology, they argue that
the distribution $F_g(x,\xi)$ at small $x$ should be 
bounded below by the forward distribution 
at $x_2=(x+\xi)/(1+\xi)$ \ , 
\begin{equation}
    2x F_g(x, \xi) \ge x_2g(x_2) \ .  
\end{equation}
Combining this with the upper bound in Eq. (\ref{gmr}), 
they concluded phenomenologically when $x<10^{-2}$ \ , 
\begin{equation}
    2xF_g(x, \xi, \mu) \sim x_2 g(x_2, \mu)\ , 
\end{equation}
at $\mu^2 = 0.4$ GeV$^2$ for the GRV gluon distribution \cite{grv} and 
at $\mu^2 = 1.3$ GeV$^2$ for the MRS(R2) distribution \cite{msr}. 

\subsection{Evolution Equations}

Parton distributions are renormalization 
scale and scheme dependent because the
defining operators are. Thus, a 
natural starting point for discussing 
scale evolution is to consider renormalization
of twist-two operators. However,
it is well known that the parton evolution
has a simple physical interpretation in 
light-cone coordinates and gauge, as
exemplified by the DGLAP evolution kernels. 
Therefore, we shall also consider 
the off-forward parton evolution kernels.

Renormalization of the twist-two
operators appearing in
Feynman parton distributions is a special
example of a more general case. 
For a given twist-two operator of spin $n$
shown in Eq. (\ref{O1}), apart from its mixing 
with the gluon operator of same quantum 
numbers it also mixes with the following
operators with total derivatives,
\begin{equation}
   {\cal O}^{\mu_1\cdots \mu_n}_{n,2i}
   = i\partial^{(\mu_1}\cdots i\partial^{\mu_{2i}}
    \overline \psi i\stackrel{\leftrightarrow}
     {\cal D}^{\mu_{2i+1}}\cdots    
   i\stackrel{\leftrightarrow}
     {\cal D}^{\mu_{n-1}} \gamma^{\mu_n)}\psi
\end{equation}
Consideration of such mixing is already 
necessary in studying the evolution of 
the leading-twist meson wave functions. The 
answer was first obtained by Efremov and Radyushkin 
\cite{efr}, and by Brodsky and Lepage \cite{bro2}. 
Actually, at the leading logarithmic order
the answer may be guessed from the naive 
conformal symmetry which is broken by quantum 
effects only at the next-to-leading-logarithmic 
order \cite{mak,ohr1}. 
The following combination of the
twist-two operators furnishes a representation
of the special conformal symmetry group,
\begin{equation}
    \tilde {\cal O}_{n} = (i\partial \cdot n)^{n-1}
     { \overline \psi} C^{3/2}_{n-1}\left({i\stackrel{
        \leftrightarrow}{\cal D}\cdot n \over i\stackrel{\leftarrow}
           {\partial} + i\stackrel{\rightarrow}
    {\partial}}\right)\not\! n \psi \ , 
\end{equation}
where the $C^{3/2}_{n-1}(x)$ are the Gegenbauer 
polynomials of order 3/2. Its evolution takes exactly 
the diagonal form of that for the
operator without the total derivatives  
\cite{efr}
\begin{equation}
   \tilde {\cal O}_{n}(\mu_1) = \left( \alpha_s(\mu_1)\over \alpha_s(\mu_2)
   \right)^{\gamma_n\over 2\beta_0}  \tilde {\cal O}_{n}(\mu_2) \ ,  
\end{equation}
where
\begin{equation}
   \gamma_n = 2C_F\left[ 4\sum_{i=1}^n{1\over i} - 3 - {2\over
n(n+1)}\right] \ , 
\end{equation}
and $\beta_0 = 11-2n_f/3$. $C_F=4/3$ for the SU(3) color
group and $n_f$ is the number of light quark flavors.
For the leading-twist pion wave 
function, the above result leads to 
a general expansion in terms of Gegenbauer polynomials.

Generalization to the singlet case is in principle 
straightforward. 
For the twist-two operators with natural parity,
the evolution was first worked out by Chase 
\cite{cha}. For those with unnatural
parity, it was worked out by Ohrndorf \cite{ohr2}, 
Shifman and Vysotsky \cite{shi}, and Baier and Grozin
\cite{bai}. The conclusion of these studies is that
the gluon towers of operators evolve in the 
leading-logarithmic approximation exactly 
like the forward case if they are combined 
according to the Gegenbauer
polynomials of order 5/2. These results were confirmed 
later in a more elegant form of light-ray or 
string operators by Geyer et al. \cite{gey},
and by Balitsky and Braun \cite{bal2}.

In light of these developments, evolution of 
off-forward distributions can simply be worked out 
by converting the anomalous
dimensions of the composite operators into 
evolution kernels. However, as in the 
case of the DGLAP evolution \cite{dglap},  
one can obtain the same results 
more directly through studying parton 
splitting processes. A first such study 
for the nonforward, helicity-independent parton 
evolution was made by Bartels and Loewe in the 
small $x$ limit \cite{bar}. Gribov et al.
presented the complete leading-order kernels 
including the singlet mixing \cite{gri}. 
M\"uller et al. found the non-singlet evolution 
kernel by generalizing the Brodsky-Lepage
result \cite{mul1}. More recently, the issue  
has been reexamined from different angles
by Radyushkin \cite{rad1,rad2},
Ji \cite{ji5}, Chen \cite{chen}, 
Frankfurt et. al. \cite{fra2}, Bl\"umlein et al. 
\cite{blu}, and Martin and Ryskin \cite{mar}.
As an example of the off-forward evolution, 
we cite below the helicity-independent result
in Ref. \cite{ji5}. 

According to Eq. (\ref{fock}), it is natural that
the distributions in three different kinematic
regions evolve differently. 
In the region $x>\xi$ where we have a quark 
creation and annihilation,
the evolution equation is,  
\begin{equation}
        { D_QF_{NS}(x,\xi,Q^2)\over D\ln Q^2} = {\alpha_s(Q^2)
             \over 2\pi}
           \int^1_x  {dy\over y} P_{NS}\left({x\over y}, 
       {\xi\over y}, {\epsilon\over y}\right) F_{NS}(y,\xi,Q^2) \ , 
\label{ap}
\end{equation}
where 
\begin{equation}
     {D_Q\over D\ln Q^2} = {d\over d\ln Q^2}
       - {\alpha_s(Q^2)\over 2\pi}C_F
      \left[{3\over 2} + \int^x_{\xi} {dy\over
        y-x-i\epsilon} + \int^x_{-\xi}{dy \over
        y-x-i\epsilon}\right] \ . 
\end{equation}
The parton splitting function is 
\begin{equation}
            P_{NS} (x,\xi,\epsilon) =C_F{x^2 + 1 - 2\xi^2 \over
             (1-x+i\epsilon)(1-\xi^2)}\ . 
\end{equation}
The end-point singularity is cancelled by the divergent integrals
in $D_Q/D\ln Q^2$. Obviously, when $\xi=0$, 
the splitting function becomes the usual DGLAP 
evolution kernel \cite{dglap}. 
For $-\xi < x < \xi$ where we have creation or annihilation
of a quark-antiquark pair, the evolution takes the form, 
\begin{equation}
        { D_QF_{NS}(x,\xi,Q^2)\over D\ln Q^2} = {\alpha_s(Q^2) \over 2\pi}
           \left[\int^1_x  {dy\over y} P'_{NS}\left({x\over y}, {\xi\over y},
       {\epsilon\over y}\right)
            - \int^x_{-1}{dy\over y} P'_{NS}\left({x\over y}, 
        -{\xi\over y}, {\epsilon\over y}\right)\right] 
        F_{NS}(y,\xi,Q^2) \ , 
\end{equation}
where 
\begin{equation}
        P'_{NS}(x, \xi, \epsilon) = 2C_F{x+\xi\over 
          \xi(1+\xi)}\left(1+{2\xi\over 1-x+i\epsilon}\right) \ . 
\end{equation}
When $\xi=1$, the kernel reduces to the Brodsky and Lepage
kernel \cite{bro2}.  For $x<-\xi$ where we have antiquark creation
and annihilation, the evolution takes 
the same form
as Eq. (\ref{ap}), apart from the replacement 
$\int^1_x \rightarrow -\int^x_{-1}$. 

The helicity-dependent off-forward distributions
were introduced in Ref. \cite{ji5} and their evolution
was also studied in that paper. Subsequently, the kernels
have been confirmed by the studies of 
Bl\"umlein et al. \cite{blu} 
and Balitsky and Radyushkin \cite{bal3}. The 
helicity-flip off-forward quark distributions were
introduced by Collins et al. \cite{col3},  
and their evolution was studied by Belitsky 
and M\"uller \cite{bel1}. The helicity-flip gluon 
distributions were defined by Hoodbhoy and 
Ji \cite{hoo2} who also derived the leading-logarithmic 
evolution. It is worthwhile to point out that the 
combinations of the helicity-flip quark and gluon operators 
according to the Gegenbauer polynomials of order 3/2 and 5/2
respectively also form representations of the naive conformal 
symmetry group and hence have diagonal evolutions 
at the leading-logarithmic level. 

\subsection{Effects of Evolution}

As in the forward case, the off-forward parton evolution
equations can be solved in many different ways. 
A few approaches have already been suggested in the
literature \cite{rad2,fra2,bel2,man1,mar}. One
important aspect of the evolution is the asymptotic
form of the off-forward distributions. Radyushkin
found \cite{rad2} (also \cite{bel2,man1}) for the  
non-singlet quark distributions,
\begin{equation}
     F_{NS}(x, \xi, t, Q^2)\stackrel{Q^2\rightarrow \infty}
     {\longrightarrow} {1\over \xi}(1-{x^2\over \xi^2})
      \theta(\xi-\xi(t))F_{NS}(t), 
\end{equation}
where $F(t)$ is the appropriate electromagnetic form factor.
This asymptotic form has interesting implications.

When $\xi=0$, the asymptotic form is a $\delta$-function
in $x$. This reminds us of the well-known fact that
as $Q^2\rightarrow \infty$, parton distributions migrate
to smaller and smaller $x$. For $\xi=1$, the evolution
is that of a meson wave function with an asymptotic
limit $1-x^2$, which has a broad shape in $x$. Therefore, 
it is quite plausible that for an intermediate 
$\xi$, one would see the migration of the distribution
from large to small $|x|$ if $|x|>\xi$. However, in the
middle region $|x|<\xi$, the total strength 
increases and at the same time, the distribution evens out 
toward $1-x^2$. These features of evolution were observed in 
the numerical calculations by Belitsky et al. \cite{bel2}
and Mankiewicz et al. \cite{man1}.

The evolution effects for the off-forward 
gluon distribution at small $x$ are especially 
interesting from the phenomenological point of
view. These effects have been studied 
by Frankfurt et al. \cite{fra2}, and Martin
and Ryskin \cite{mar}. In Ref. \cite{fra2}, 
they assumed negligible off-diagonal effects 
at $Q_0 = 1.6$ GeV, and then evolved 
the off-diagonal distribution to $Q = 7$, 17, 41, and 110 GeV 
at $x_1=1.1\times 10^{-2}, 1.1\times 10^{-3}$, 
and $1.1\times 10^{-4}$. For $x_2/x_1 = 10^{-2}$ to 1, they
found that the ratio of off-diagonal to diagonal 
distributions ranges from 1.2 to 1.7. They pointed out that
for a fixed $x_2-x_1$, the off-diagonal 
distribution at large $Q$ is obtained from that in the
much larger $x$ region, however, with the same difference
$x_2-x_1$ where the off-diagonal effect is very small. 
Therefore at very small $x$, unless $Q$ is very
close to hadron mass scales, the off-diagonal 
effects come mostly from the perturbative evolution. 
These effects are less than a factor
of 2 for the kinematics region considered.

Similar studies have also been made in Ref. \cite{mar}.
The qualitative conclusion is the same although 
quantitatively there are differences. Martin and
Ryskin pointed out that the off-diagonal evolution 
happens faster than the diagonal one because the 
difference in the splitting functions is positive. 
Furthermore, the difference is only important  
when $Q$ is close to the final target scale
and $x$ becomes comparable to the size
of $\xi$. 

\section{Probing OFPD's in High-energy Experiment}

Eventually, usefulness of the OFPD's depends on 
whether they can actually be measured in any 
experiment. The simplest, and possibly  
the most promising, type of experiments 
are hard electro- or deep-inelastic 
production of photon, mesons, and perhaps even 
lepton pairs. In this section, we will consider
separately two experiments: deeply virtual 
Compton scattering (DVCS) in which a real photon 
is produced, and diffractive
meson production. There are practical advantages
and disadvantages from both processes. Real
photon production is, in a sense, cleaner but the
cross section is reduced by an additional power
of $\alpha_{\rm em}$. Moreover, the DVCS amplitude  
interferes with the Bethe-Heitler one. 
Meson production may be easier to detect, however, 
it has a twist suppression, $1/Q^2$. 
In addition, the theoretical cross section 
depends on the unknown light-cone meson wave 
function. At present, there exist already some 
data from HERA on meson production
\cite{cri}. More 
experiments will be done in the future at HERA, 
CERN, and Jefferson Lab. 

\subsection{Deeply Virtual Compton Scattering}

Deeply virtual Compton scattering was proposed 
in Ref. \cite{ji1} as a practical way to measure 
the off-forward distributions. The result was 
confirmed in Ref. \cite{rad3} using the  
double distributions.
The general two-photon processes had actually 
been studied earlier by Watanabe \cite{wat} 
and M\"uller et al. \cite{mul1}, and later 
by Chen \cite{chen}. Recently, there have been
a number of new results obtained about 
this process, which will be described below. 
 
Consider the virtual photon scattering shown in Fig. 2a,
where the momenta of the incoming (outgoing)
photon and nucleon are $q(q')$ and $P(P')$,
respectively. The Compton amplitude is defined
as 
\begin{equation}
    T^{\mu\nu} = i\int d^4z e^{\bar q\cdot z}
    \left\langle P'\left|{\rm T}J^\mu\left(-{z\over 2}\right)
    J^\nu\left({z\over 2}\right)\right|P\right\rangle
\end{equation}
where $\bar q = (q+q')/2$. In the Bjorken 
limit, $-q^2 $ and $ P\cdot q\rightarrow 
\infty$ and their ratio remains finite,
the scattering is dominated by the single quark 
process shown in Fig. 2b in which a quark absorbs 
the virtual photon, immediately radiates a real one, 
and falls back to form the recoiled nucleon. 
In the process, the initial and final photon 
helicities remain the same. The leading-order 
Compton amplitude is \cite{ji1} 
\begin{eqnarray}
     T^{\mu\nu} &=& g^{\mu\nu}_\perp \int^1_{-1}
       dx \left({1\over x-\xi+ i\epsilon} 
       + {1\over x+\xi-i\epsilon}\right) \sum_q 
       e_q^2 F_q(x, \xi, t, Q^2) 
    \nonumber \\  
     && + ~ i \epsilon^{\mu\nu\alpha\beta} p_\alpha n_\beta
      \int^1_{-1}
      dx  \left({1\over x-\xi + i\epsilon} 
       - {1\over x+\xi-i\epsilon}\right) \sum_q e_q^2 
      \tilde F_q(x, \xi, t, Q^2) \ ,      
\end{eqnarray}
where $n$ and $p$ are the conjugate light-cone vectors
defined according to the collinear direction of  
$\bar q$ and $\bar P$, and $g^{\mu\nu}_\perp$
is the metric tensor in the transverse space. $\xi$ is related
to the Bjorken variable $x_B = - q^2 /(2 P\cdot q)$
by $x_B=2\xi/(1+\xi)$. 

\begin{figure}
\label{fig2}
\epsfig{figure=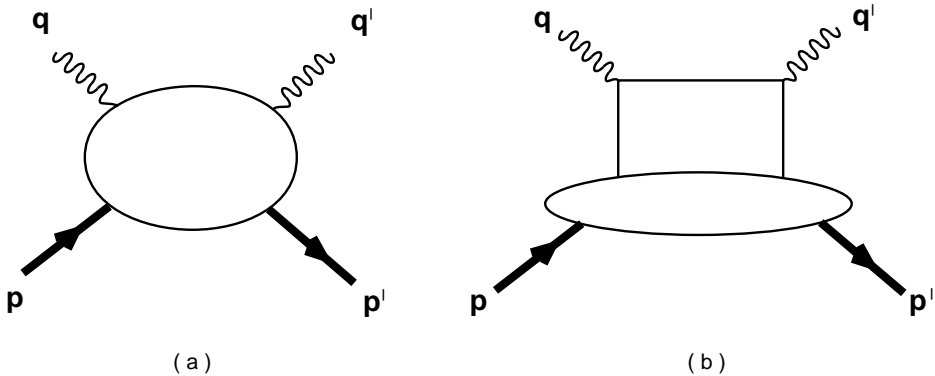,height=5.0cm}
\caption{a). Compton scattering; b) Leading Feynman 
diagrams for DVCS.}
\end{figure}  

The same DVCS final state can also be produced
through the Bethe-Heitler process in which the initial
or final state electron (or muon) radiates a
real photon and at the same time scatters elastically
off the target nucleon.  This process can be calculated 
as accurately as the data on the nucleon 
elastic form factors. Physically, the Bethe-Heitler 
process is a background to DVCS and can overshadow the
latter signal if the former is too large. On the 
other hand, with an appropriate size of 
Bethe-Heitler amplitude, one may hope to measure its 
interference with the DVCS amplitude which then can be
directly extracted from data. The 
complete leading-order DVCS cross sections and
the interference with the Bethe-Heitler amplitude
can be found in Ref. \cite{ji5}. 

Recently, Guichon made some interesting estimates of
the cross sections at COMPASS and Jefferson Lab
energies \cite{gui}. He found that while at low scattering
energy, the Bethe-Heitler background is 
quite large, it drops significantly at high energies
when the virtual photon flux is large. In fact, 
if one is focused on production of real photons 
along the direction of the virtual 
photon, the DVCS cross section is quite
significant at COMPASS energy, and certainly far
above the Bethe-Heitler background. As far as 
studying the orbital angular momentum of quarks
is concerned, this is the most interesting 
kinematic region. Guichon's estimate
is encouraging for a real experiment.

The one-loop corrections to DVCS have 
recently been studied by Osborne and 
Ji\cite{ji6}. They have also been studied by 
M\"uller \cite{mul2} and by Belitsky and 
M\"uller \cite{bel1} using the constraints
from conformal symmetry. 
Results of these studies have been confirmed by
Mankiewicz et al. \cite{man2}. Besides
their obvious use in precision analysis, 
the results indicate that DVCS is factorizable at the 
next-to-leading order. A higher-order calculation of 
the coefficient functions in the renormalon-chain 
approximation has been done by Belitsky and 
Sch\"afer \cite{bel3}. An all order proof of the 
DVCS factorization  was first given by Radyushkin \cite{rad2} 
and recently in a different perspective 
by Osborne and Ji \cite{ji7}, and Collins and Freund \cite{col4}. 

The simple DVCS scattering mechanism
can be tested through scaling relations 
and selection rules, as in the case of
deep-inelastic scattering.
In Ref. \cite{die}, Diehl et al. considered
the possibility of using $\phi$ dependence of the
cross section to test the single quark scattering
picture, where $\phi$ is the angle between the lepton
and hadron scattering planes.
They observed that through QCD radiative corrections,
the photon helicity-flip amplitude may
contribute to DVCS. A recent study by Hoodbhoy and
Ji showed that this indeed happens \cite{hoo2}. 
The key element here is the helicity-flip 
off-forward gluon distributions, which decouple 
in the forward limit due to angular momentum 
conservation. They pointed out that a nonvanishing 
leading-twist photon helicity-flip amplitude 
signals unequivocally the presence of 
gluons in the nucleon. 

There are other processes that are as clean
as the DVCS, but more difficult to measure 
experimentally. For instance, $Z^0$ production was 
considered in Ref. \cite{bar}. Since $Z^0$'s are 
signaled through lepton pairs, one can 
also consider lepton-pair production 
through intermediate time-like 
virtual photons. Theoretical treatment of these
is exactly the same as for DVCS. 

\subsection{Vector Meson Production}

Heavy quarkonium production was first considered 
by Ryskin as a way to measure the Feynman gluon
distributions at small $x$ \cite{rys1}. 
A leading-order diagram for the process
is shown in Fig 3a, where the virtual photon 
fluctuates into a $c\bar c$ pair which subsequently
scatters off the nucleon target through two-gluon 
exchanges. In the process, the pair transfers
away certain amount of its longitudinal momentum and
reduces its invariant mass to that of a $J/\Psi$. 
The cross section is,
\begin{equation}
    {d\sigma \over dt}(\gamma^* +P\rightarrow
       J/\Psi+P') = {16\pi^3M\Gamma_{e^+e^-} \over 3\alpha_{\rm em}Q^6}
       \alpha_s^2(\bar Q^2)[\xi g(\xi, \bar Q^2)]^2
\end{equation}
where $\bar Q^2 = (Q^2+M^2)/4$, $M$ is the 
$J/\psi$ mass, and $\Gamma_{e^+e^-}$ is the decay
width into the lepton pair. The equation was derived
in the kinematic limit $s\gg Q^2\gg M^2\gg t$
and the Fermi motion of the quarks in the 
meson was neglected. Two other important
approximations were used in the derivation.
First, the contribution from the real part
of the amplitude is neglected, which may be
justifiable at small $x$. Second, the 
off-forward distributions are identified with the 
forward ones. As we have discussed in IV.A, even
for small $t$, the approximation is good only 
if the off-diagonal evolution effects can 
be neglected. A more refined treatment of the 
process and comparison with HERA data can be found in 
Ref. \cite{rys2}. 

\begin{figure}
\label{fig3}
\epsfig{figure=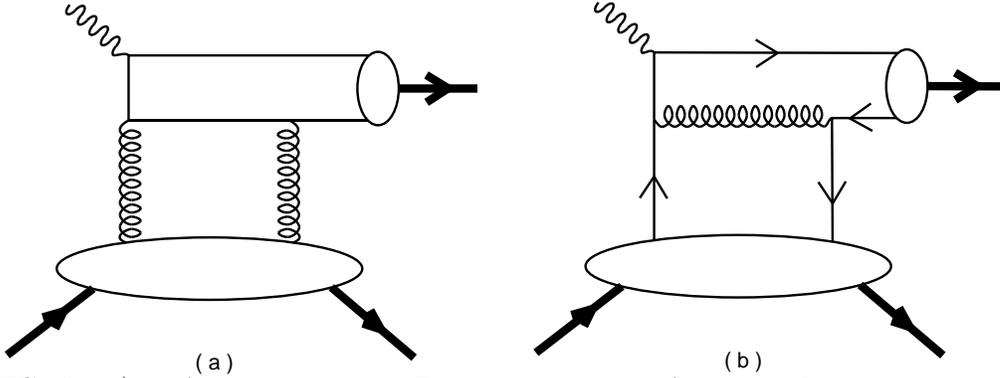,height=5.0cm}
\caption{a). A representative Feynman diagram
for neutral meson production through off-forward 
gluon distributions in the nucleon; 
b) A representative Feynman diagram 
for meson production through off-forward
quark distributions in the nucleon.} 
\end{figure}  

The above result was extended to the case
of light vector-meson production by 
Brodsky et al., who considered the effects
of meson structure in perturbative QCD \cite{bro1}. 
They found a similar cross section,
\begin{equation}
  \left. {d\sigma \over dt}\right|_{t=0}
     (\gamma^*N\rightarrow VN) 
   = {4\pi^3\Gamma_V m_V \alpha_s^2(Q) 
   \eta_V^2\left(xg(x, Q^2)\right)^2 \over 
    3\alpha_{\rm em} Q^6} \ , 
\end{equation}
where the dependence on the meson structure 
is in the parameter
\begin{equation}
\eta_V = {1\over 2} \int {dz\over z(1-z)} \phi^V(z)
     \left( \int dz \phi^V(z) \right)^{-1} \ ,  
\end{equation}
and $\phi^V(z)$ is the leading-twist light-cone 
wave function. Evidently, the above formula 
reduces to Ryskin's in the heavy-quark limit
$\left( \phi^V(x) = \delta(x-1/2)\right)$. More 
discussions on vector meson production along this 
line, including comparison with recent data, can be found
in Ref. \cite{rys2,fra1,rys3}. 

Radyushkin was the first to calculate the amplitude
of hard diffractive electroproduction in terms
of off-forward gluon distributions \cite{rad1,rad2}. With 
the virtual photon and vector meson both polarized longitudinally, 
his result corresponds to
\begin{equation}
 {d\sigma \over dt}
     (\gamma^*N\rightarrow VN) 
   = {4\pi\Gamma_V m_V \alpha_s^2(Q) 
   \eta_V^2 \over  3\alpha_{\rm em} Q^6}
     \left|2x_B\int^1_{-1}dx
      \left({1\over x-\xi+i\epsilon} + 
        {1\over x+\xi-i\epsilon}\right)F_g(x,\xi,t)\right|^2 \ , 
\end{equation}
where again $x_B = 2\xi/(1+\xi)$. The above formula
is valid for any $x_B$ and $t$ smaller or around
hadron mass scales. Hoodbhoy has also studied the effects of 
the off-forward distributions in the case of
$J/\psi$ production \cite{hoo1}. He found
that Ryskin's result needs to be modified 
in a similar way once the off-forward effects
becomes important. 

Collins, Frankfurt, and Strikman have made
an extensive analysis of the high-order 
QCD contributions to hard diffractive meson
production \cite{col3}. Using the standard
tools described in \cite{col2}, they
showed that the leading-twist contributions 
are factorizable to all orders in perturbation
theory. The factorization theorem provides
a rigorous QCD basis for calculating the process
in terms of the off-forward distributions. 
In fact, the paper outlined a recipe for
calculating a whole class of processes
in perturbation theory. Mankiewicz, 
Piller and Weigl recently calculated the
neutral pseudoscalar and vector meson production cross 
sections in terms of the non-forward quark 
and gluon distributions \cite{man2}. Shown in 
Fig. 3b is a Feynman diagram in which 
meson production depends on a quark distribution.
They also investigated charged particle
production in which the initial nucleon has
a different charge from the final one \cite{man3}. 
The charge-changing off-forward distributions 
can be obtained from the charge-conserving 
ones using isospin symmetry. The result 
can be generalized to the octet baryons 
if the SU(3) flavor symmetry is good. One can also 
consider production of heavy-flavored baryons from 
the nucleon target. Unlike the heavy-quark
distributions in light hadrons, the 
corresponding off-forward distributions are not suppressed
by heavy-quark masses.

\section{summary and outlook}

The off-forward parton distributions are a new type
of nucleon observables which have recently generated 
much interest in hadron physics community. 
On the one hand, the OFPD's contain all the 
form factors of the twist-two operators 
which generalize the well-known electroweak 
currents. The form factors of the spin-two, 
twist-two operators (energy-momentum tensor)
are especially interesting in unravelling the spin
structure of the nucleon. On the other hand, they
summarize succinctly the leading structural 
dependence in a class of high-energy processes 
involving the nucleon. The partonic
interpretation of the OFPD's is quite simple--
it represents parton-nucleon scattering 
amplitudes on the light-cone.

Recent studies of the OFPD's can largely be
classified into three categories: modelling the 
new distributions, studying their scale dependences,
and investigating processes that are sensitive 
to them. In terms of model studies, our accomplishments
so far are rather limited. Even though there are bag 
and chiral soliton model calculations, these
calculations need much improvement 
in the future. In particular, the physics in 
the $|x|<\xi$ region has not been entirely 
captured in these calculations. The 
models are yet to be extended to including
gluon degrees of freedom which are 
known to be important even at low energy scales. 
Indeed, a vanishing helicity-flip gluon 
distribution at low-energy scales will remain  
zero at other scales because it has no mixing 
with any quark distribution. A number of rigorous 
constraints discussed in Section IV.A can be helpful
in modelling: Apart from the special limits
and $n=1$ moments, there exist useful upper bounds
for the OFPD's.
  
There have been many studies in perturbative 
evolution. At the leading-logarithmic level, evolution
equations now have been obtained for all leading-twist
OFPD's. At next-to-leading order, one
needs two-loop anomalous dimensions which 
exist only for the spin-independent non-singlet
operators \cite{tl}. Calculations of additional two-loop
anomalous dimensions will be needed as 
data on the OFPD's become available. Recent
works of M\"uller and Belitsky illustrate 
the power of exploiting the anomalous 
conformal symmetry to obtain these anomalous
dimensions \cite{b1}.  
Numerical studies of evolution effects
have just begun. Qualitatively we can understand
the evolution effects from its asymptotic 
behavior. Quantitatively, one needs to 
establish high-precision, high-speed evolution
codes which allow efficient data fitting 
in the future.

The leading-order cross sections for DVCS and
meson production can be found in the literature. 
For DVCS, the cross section has also been
computed to the next-to-leading order in $\alpha_s$. 
A similar study for meson production can be made,
and will be necessary once precision
data become available. Higher-twist effects
are important at low $Q^2$ and must be

investigated thoroughly. Generally speaking, they are 
more complex to study than that in 
deep-inelastic scattering because of the extra
photon or meson, and the extra soft scale $t$. 
Additional processes 
may be explored to probe the OFPD's, including
ones in which the final state baryon may contain
strange or heavy flavors. Polarization 
observables need to be seriously
investigated. A solid experimental confirmation
of the scattering mechanisms involving the OFPD's
is itself very significant. It involves
testing scaling relations and selection rules.

Finally, we notice that the OFPD's are functions
of multi-variables. As such, it would be 
difficult to get a complete picture of them 
in a short time period and in a single type of experiments. 
Moreover, experimental cross sections cannot be 
used to extract the distributions directly; 
they usually involve integrals over parton 
momenta. Therefore, it will be necessary to 
parametrize the OFPD's and fit the parameters 
to data. In finding appropriate ways to make
parametrizations and fits, model studies
will play important role, unless of course
lattice calculations can make a significant breakthrough
in the near future.

{\bf Note Added:} After this paper was submitted for 
publication, the author has noticed some new works in the 
literature which are relevant to the subject of this paper. 
In Ref. \cite{rad11}, Radyushkin studied both hadron
form factors and wide-angle Compton scattering
using nonforward parton densities. In Ref. 
\cite{gol}, diffractive dijet photoproduction
was proposed as a probe of the off-diagonal
gluon distribution. In Ref. \cite{pir}, certain
positivity constraints for off-forward
parton distributions were derived. In a different
line of development, the orbital angular momentum 
distribution in the nucleon was investigated
\cite{orb}.  

\acknowledgements
In the last two years I have benefited considerably  
from discussions surrounding the topic of this
article with S. Brodsky, J. Collins, P. Hoodbhoy, 
A. Mueller, J. Osborne, A. Radyushkin, G. Sterman, 
M. Strikman, and others, and I wish to thank them all.  
I also would like to thank A. Martin for his 
encouragement in writing this topical review, K. Golec-Biernat, 
A. Radyushkin and L. Robinette for a critical 
reading of the manuscript, and J. Osborne for drawing the figures. 
This work is supported in part by funds provided by the U.S. 
Department of Energy (D.O.E.) under cooperative agreement 
DOE-FG02-93ER-40762.


\begin{references}
\frenchspacing

\bibitem{fey}
R. P. Feynman, Phys. Rev. Lett. {\bf 23} (1969) 1415; {\it 
Photon Hadron Interactions}, Benjamin, New York, 1972. 

\bibitem{col1}
J. C. Collins, G. Sterman, and D. E. Soper, in {\it Perturbative
Quantum Chromodynamcis}, ed. A. Mueller (World Scientific, 
Singapore, 1989)

\bibitem{col2}
J. C. Collins and D. E. Soper, Nucl. Phys. {\bf B194} (1982) 445. 

\bibitem{lat}
See for instance, M. Creutz, {\it Quarks, Gluons and Lattices}
(Cambridge Univ. Press, Cambridge, U.K., 1983) 

\bibitem{grv}
M. Gl\"uck, E. Reya, and A. Vogt, Z. Phys. {\bf C 67} (1995) 433.

\bibitem{msr}
A. D. Martin, R. G. Roberts, and W. J. Stirling,
Phys. Lett. {\bf B387} (1996) 419. 

\bibitem{cteq}
H. L. Lai et al., Phys. Rev. {\bf D55} (1997) 1280. 

\bibitem{wat}
K. Watanabe, Prog. Theo. Phys. {\bf 67} (1982) 1834. 

\bibitem{bar}
J. Bartels and M. Loewe, Z. Phys. {\bf C12} (1982) 263.

\bibitem{gri}
L. V. Gribov, E. M. Levin, and M. G. Ryskin, 
Phys. Rep. {\bf 100} (1983) 1. 

\bibitem{rys1}
M. G. Ryskin, Z. Phys. {\bf C37} (1993) 89. 

\bibitem{bro1}
S. J. Brodsky, L. L. Frankfurt, J. F. Gunion, A. H. Mueller, 
and M. Strikman, Phys. Rev. {\bf D50} (1994) 3134.

\bibitem{rys2}
M. G. Ryskin, R. G. Roberts, A. D. Martin, and E. M. Levin,
Zeit. Phys. C76 (1997) 231. 

\bibitem{fra1}
L. L. Frankfurt, W. Koepf, and M. Strikman, Phys. Rev. {\bf D54} 
(1996) 3194.

\bibitem{dit}
F. M. Dittes, D. Muller, D. Robaschik, B. Geyer, and J. Horejsi,
Phys. Lett. {\bf B 209} (1988) 325. 

\bibitem{dglap}
V. N. Gribov and L. N. Lipatov, Sov. J. Nucl. Phys. {\bf 15}
(1972) 78; G. Altarelli and G. Parisi, Nucl. Phys. {\bf B126}
(1977) 298; Yu. L. Dokshitser, Sov. Phys. JETP {\bf 46} (1977)
641.

\bibitem{efr}
A. V. Efremov and A. V. Radyushkin, JINR-E2-11535, Dubna, 1978;
A. V. Efremov and A. V. Radyushkin, Phys. Lett. {\bf B94} (1980) 245. 

\bibitem{bro3}
S. J. Brodsky and G. P. Lepage, Phys. Rev. {\bf D22} (1980) 2157. 

\bibitem{mul1}
D. M\"uller, D. Robaschik, B. Geyer, F.-M. Dittes, 
and J. Horejsi, Fortschr. Phys. {\bf 42} (1994) 101.

\bibitem{jai}
P. Jain and J. P. Ralston, in the proceedings of the workshop
on Future Directions in Particle and Nuclear Physics at Multi-GeV
Hadron Beam Facilities, BNL, March, 1993.

\bibitem{ji1}
X. Ji, Phys. Rev. Lett. {\bf 78} (1997) 610. 

\bibitem{abr}
H. Abramowicz, L. Frankfurt, and M. Strikman, DESY-95-047, 
SLAC Summer Inst. 1994, 539. 

\bibitem{rad1}
A. V. Radyushkin, Phys. Lett. {\bf B385} (1996) 333. 

\bibitem{hoo1}
P. Hoodbhoy, Phys. Rev. {\bf D56} (1997) 388.

\bibitem{col3}
J. C. Collins, L. Frankfurt, and M. Strikman, 
Phys. Rev. {\bf D56} (1997) 2982. 

\bibitem{iz}
C. Itzykson and J. Zuber, {\it Quantum Field Theory}, 
McGraw-Hill Inc., New York, 1980. 

\bibitem{bro2}
S. Brodsky and P. Lepage, in {\it Perturbative
Quantum Chromodynamcis}, ed. A. Mueller (World Scientific, 
Singapore, 1989); See also X. Ji, Comm. Nucl. Part. 
Phys. {\bf 21} (1993) 123. 

\bibitem{hoo2}
P. Hoodbhoy and X. Ji, hep-ph/9801369. 

\bibitem{jaf1}
R. L. Jaffe and X. Ji, Nucl. Phys. {\bf B375} (1992) 527. 

\bibitem{rad2}
A. V. Radyushkin, Phys. Rev. {\bf D56} (1997) 5524. 

\bibitem{rad3}
A. V. Radyushkin, Phys. Lett. {\bf B380} (1996) 417.

\bibitem{alt1}
G. Alteralli, R. D. Ball, S. Forte, and G. Ridolfi,
Nucl. Phys. {\bf B496} (1997) 337. 

\bibitem{fuk}
M. Fukugita, Y. Kuramashi, M. Okawa, and A. Ukawa, 
Phys. Lett. {\bf 75} (1995) 2092.

\bibitem{don}
S. J. Dong, J. F. Lagae, K. F. Liu, Phys. Rev. Lett. {\bf 75} (1995) 2096.

\bibitem{goc}
M. G\"ockeler et al., Phys. Rev. {\bf D53} (1996) 2317. 

\bibitem{jaf2}
R. L. Jaffe and A. Manohar, Nucl. Phys. {\bf B337} (1990) 509.

\bibitem{seh}
L. Sehgal, Phys. Rev. {\bf D10} (1974) 1663. 

\bibitem{rat}
P. G. Ratcliffe, Phys. Lett. {\bf B192} (1987) 180. 

\bibitem{alt2}
G. Altarelli and and G. G. Ross, Phys. Lett. {\bf B212} (1988) 391.

\bibitem{car}
R. D. Carlitz, J. C. Collins, and A. H. Mueller, 
Phys. Lett. {\bf B214} (1988) 229.

\bibitem{ji2}
X. Ji, hep-ph/9710290.

\bibitem{ji3}
X. Ji, J. Tang, and P. Hoodbhoy, 
Phys. Rev. Lett. {\bf 76} (1996) 740. 

\bibitem{gro}
D. Gross and F. Wilczek, Phys. Rev. {\bf D9} (1974) 980. 

\bibitem{bal1}
I. I. Balitsky and X. Ji, Phys. Rev. Lett. {\bf 79} (1997) 1225.

\bibitem{baro}
V. Barone, T. Calarco, A. Drago, hep-ph/9801281. 
 
\bibitem{mar}
A. Martin and M. G. Ryskin, hep-ph/9711371, Phys. Rev. D57, June
1st 1998. 

\bibitem{ji4}
X. Ji, W. Melnitchouk, and X. Song, {\bf D56} (1997) 5511.

\bibitem{pet}
V. Yu. Petrov, P. V. Pobylitsa, M. V. Polyakov, 
I. B\"ornig, K. Goeke, and C. Weiss, hep/ph9710270. 

\bibitem{sig}
A. W. Schreiber, A. I. Signal, A. W. Thomas, Phys. Rev. {\bf D44}
(1991) 2653.  

\bibitem{mak}
Yu. M. Makeenko, Sov. J. Nucl. Phys. {\bf 33} (1982) 440. 

\bibitem{ohr1}
Th. Ohrndorf, Nucl. Phys. {\bf B198} (1982) 26. 

\bibitem{cha}
M. K. Chase, Nucl. Phys. {\bf B174} (1980) 109; 

\bibitem{ohr2}
Th. Ohrndorf, Nucl. Phys. {\bf B186} (1981) 153; 

\bibitem{shi}
M. A. Shifman and M. Vysotsky, Nucl. Phys. {\bf B186} 
(1981) 475. 

\bibitem{bai}
V. N. Baier and A. G. Grozin, Nucl. Phys. {\bf B192} (1981) 476. 

\bibitem{gey}
B. Geyer, D. Robaschik, M. Bordag, and J. Horejsi, Z. Phys. 
{\bf C26} (1985) 591;
I. Braunschweig, B. Geyer, J. Horejsi and D. Robaschik, Z. Phys. 
{\bf C33} (1987) 175. 

\bibitem{bal2}
I. I. Balitsky and V. M. Braun, Nucl. Phys. {\bf B311} (1988) 541. 

\bibitem{ji5}
X. Ji, Phys. Rev. {\bf D55} (1997) 7114. 

\bibitem{chen}
Z. Chen, hep-ph/9705279. 

\bibitem{fra2}
L. L. Frankfurt, A. Freund, V. Guzey, and M. Strikman, hep-ph/9703449.

\bibitem{blu}
J. Bl\"umlein, B. Geyer, and D. Robaschik, Phys. Lett. {\bf B413} 
(1997) 114. 

\bibitem{bal3}
I. I. Balitsky and A. V. Radyushkin,
Phys. Lett. {\bf B413} (1997) 114. 

\bibitem{bel1}
A. V. Belitsky and D. M\"uller, hep-ph/9709379. 

\bibitem{bel2}
A. V. Belitsky, B. Geyer, D. M\"uller, and A. Sch\"afer, 
hep-ph/9710427. 

\bibitem{man1}
L. Mankiewicz, G. Piller, and T. Weigl, hep-ph/9711227.

\bibitem{cri}
J. Crittenden, hep-ex/9709031. 

\bibitem{gui}
P.A.M. Guichon, DAPNIA-SPHN-97-31, Talk given at 5th
International Workshop on Deep Inelastic Scattering 
and QCD (DIS 97), Chicago, IL, 14-18 Apr 1997. 

\bibitem{ji6}
X. Ji and J. Osborne, Phys. Rev. {\bf D57} (1998) R1337.

\bibitem{mul2}
D. M\"uller, hep-ph9704406, CERN-TH/97-80, 1997. 

\bibitem{man2}
L. Mankiewicz, G. Piller, E. Stein, M. V\"anttinen,
T. Weigl, hep-ph/9712251. 

\bibitem{bel3}
A. V. Belitsky and A. Sch\"afer, hep-ph/9801252.

\bibitem{ji7}
X. Ji and J. Osborne, hep-ph/9801260.

\bibitem{col4}
J. C. Collins and A. Freund, hep-ph/9801262.

\bibitem{die}
M. Diehl, T. Gousset, B. Pire, and J. P. Ralston, 
Phys. Lett. {\bf B411} (1997) 193. 

\bibitem{rys3}
A. D. Martin, M. G. Ryskin, and T. Teubner, 
Phys.Rev. D55 (1997) 4239; D56 (1997) 3007. 

\bibitem{man3}
L. Mankiewicz, G. Piller and T. Weigl, hep-ph/9712508. 

\bibitem{tl}
F.-M. Dittes and A. V. Radyushkin, Phys. Lett. {\bf 134B} (1984) 359;
M. H. Sarmadi, {\it ibid.} {\bf 143B} (1984) 471; G. R. Katz, 
Phys. Rev. {\bf D31} (1985) 652; S. V. Mikhailov and A. V.
Radyushkin, Nucl. Phys. {\bf B254} (1985) 89.

\bibitem{b1}
D. M\"uller, Phys. Rev. {\bf D49} (1994) 2525; 
A. V. Belitsky and D. M\"uller, hep-ph/9802411.

\bibitem{rad11}
A. V. Radyushkin, hep-ph/9803316. 

\bibitem{gol}
K. Golec-Biernat, J. Kwiecinski, and A. D. Martin, 
hep-ph/9803464.

\bibitem{pir}
B. Pire, J. Soffer, and O. Teryaev, hep-ph/9804284.
 
\bibitem{orb}
P. H\"agler and A. Sch\"afer, hep-ph/9802362; 
A. Harindranath and R. Kundu, hep-ph/9802406;
O. V. Teryaev, hep-ph/9803403; 
P. Hoodbhoy, X. Ji, and W. Lu, hep-ph/9804337. 
\nonfrenchspacing
\end{references}
\end{document}